\documentclass[zpreprint,zbstdefault,british]{zeus_K2.9paper}
\usepackage{zeusdet_units}
\chardef\usc=95
\chardef\til=126
\catcode`\@=11 % @ signs are now treated as letters
\DeclareRobustCommand\xdotspace{\futurelet\@let@token\@xdotspace}
\def\@xdotspace{%
  \ifx\@let@token.\else
  \ifx\@let@token\bgroup.\else
  \ifx\@let@token\egroup.\else
  \ifx\@let@token\/.\else
  \ifx\@let@token\ .\else
  \ifx\@let@token~.\else
  \ifx\@let@token!.\else
  \ifx\@let@token,.\else
  \ifx\@let@token:.\else
  \ifx\@let@token;.\else
  \ifx\@let@token?.\else
  \ifx\@let@token/.\else
  \ifx\@let@token'.\else
  \ifx\@let@token).\else
  \ifx\@let@token-.\else
  \ifx\@let@token\@xobeysp.\else
  \ifx\@let@token\space.\else
  \ifx\@let@token\@sptoken.\else
   .\space
   \fi\fi\fi\fi\fi\fi\fi\fi\fi\fi\fi\fi\fi\fi\fi\fi\fi\fi}
\catcode`\@=12 % @ signs are no longer letters

\newcommand{\stru}[2]{%
   \relax\ifmmode\hbox{\vrule height#1 depth#2 width0pt}%
   \else\vrule height#1 depth#2 width0pt\fi}
\newcommand{\Ronum}[1]{\uppercase\expandafter{\romannumeral#1}}
\newcommand{\ronum}[1]{\expandafter{\romannumeral#1}}
\DeclareRobustCommand{\LaTeXZ}{%
  \LaTeX\kern-.05em4\kern-.1em
  {\raisebox{-0.2ex}{$\scriptstyle\text{ZEUS}$}}\xspace}
\newcommand{\slashfrac}[2]{%
  \raisebox{0.5ex}{\ensuremath #1}\kern-0.12em/\kern-0.08em
  \raisebox{-.8ex}{\ensuremath #2}}
\newcommand{\sqr}[3]{%
    {\vcenter{\hrule height.#3ex\hbox{\vrule width.#2ex height#1ex
     \kern#1ex\vrule width.#3ex}\hrule height.#2ex}}}

\catcode`\@=11 % @ signs are now treated as letters
\newcommand{\parenbar}{\mathpalette\p@renb@r}
\def\p@renb@r#1#2{\vbox{%
  \ifx#1\scriptscriptstyle \dimen@.7em\dimen@ii.2em\else
  \ifx#1\scriptstyle \dimen@.8em\dimen@ii.25em\else
  \dimen@1em\dimen@ii.4em\fi\fi \offinterlineskip
  \ialign{\hfill##\hfill\cr
    \vbox{\hrule width\dimen@ii}\cr
    \noalign{\vskip-.3ex}%
    \hbox to\dimen@{$\mathchar300\hfil\mathchar301$}\cr
    \noalign{\vskip-.3ex}%
    $#1#2$\cr}}}
\catcode`\@=12 % @ signs are no longer letters

\ifzeusunit
  
\else
  
\fi

\newcommand{\IP}{{\rm I$\kern-0.01667em$P}\xspace}

\mathchardef\qsm=63
\mathchardef\pls=43
\mathchardef\mns=512
\mathchardef\plm=518
\mathchardef\eql=61
\mathchardef\smallleft=300
\mathchardef\smallright=301
\mathchardef\les=316
\mathchardef\gre=318
\mathchardef\leq=532
\mathchardef\grq=533
\catcode`\@=11 % @ signs are now treated as letters
\newcounter{pict@width}
\newcounter{pict@height}
\newlength{\pict@scale}
\newcommand{\psfigadd}[4]{%
\setcounter{pict@width}{1*\ratio{#2+\pict@scale/2}{\pict@scale}}
\setcounter{pict@height}{1*\ratio{#3+\pict@scale/2}{\pict@scale}}
\setlength{\unitlength}{\pict@scale}
\hbox to #2{\hspace{-\fill}\begin{picture}(\thepict@width,\thepict@height)
\put(0,0){\psfig{figure=#1,width=#2,height=#3,clip=}}
\SetScale{0.283466457}
\SetWidth{1.763889}
{#4}
\end{picture}}
}
\newcounter{pict@widthfst}
\newcounter{pict@widthscd}
\newcounter{pict@widthtot}
\newcommand{\psfigaddtwo}[7]{%
\setcounter{pict@widthfst}{1*\ratio{#2+\pict@scale/2}{\pict@scale}}
\setcounter{pict@widthscd}{1*\ratio{#2+#4+\pict@scale/2}{\pict@scale}}
\setcounter{pict@widthtot}{1*\ratio{#2+#4+#6+\pict@scale/2}{\pict@scale}}
\setcounter{pict@height}{1*\ratio{#3+\pict@scale/2}{\pict@scale}}
\setlength{\unitlength}{\pict@scale}
\hbox{\hspace{-\fill}\begin{picture}(\thepict@widthtot,\thepict@height)
\put(0,0){\psfig{figure=#1,width=#2,height=#3,clip=}}
\put(\thepict@widthscd,0){\psfig{figure=#5,width=#6,height=#3,clip=}}
\SetScale{0.283466457}
\SetWidth{1.763889}
{#7}
\end{picture}}
}
\newcommand{\psfigror}[4]{%
\setcounter{pict@width}{1*\ratio{#2+\pict@scale/2}{\pict@scale}}
\setcounter{pict@height}{1*\ratio{#3+\pict@scale/2}{\pict@scale}}
\setlength{\unitlength}{\pict@scale}
\hbox{\begin{picture}(\thepict@width,\thepict@height)
\put(0,\thepict@height){\psfig{figure=#1,width=#3,height=#2,clip=,angle=270}}
\SetScale{0.283466457}
\SetWidth{1.763889}
{#4}
\end{picture}}
}
\newcommand{\psfigrol}[4]{%
\setcounter{pict@width}{1*\ratio{#2+\pict@scale/2}{\pict@scale}}
\setcounter{pict@height}{1*\ratio{#3+\pict@scale/2}{\pict@scale}}
\setlength{\unitlength}{\pict@scale}
\hbox{\begin{picture}(\thepict@width,\thepict@height)
\put(0,0){\psfig{figure=#1,width=#3,height=#2,clip=,angle=90}}
\SetScale{0.283466457}
\SetWidth{1.763889}
{#4}
\end{picture}}
}
\catcode`\@=12 % @ signs are no longer letters
\newlength\listtextwidth

\catcode`\@=11 % @ signs are now treated as letters
\newlength{\@tabfninsert}
\newlength{\@tabfnwidth}
\newcommand{\tabfootnote}[2]{%
  \setlength{\@tabfninsert}{0.8em}
  \setlength{\@tabfnwidth}{\textwidth}
  \addtolength{\@tabfnwidth}{-\@tabfninsert}
  \addtolength{\@tabfnwidth}{-0.4em}
  \noindent\makebox[\@tabfninsert][r]{\footnotesize$^{#1}$\hfil}\hfill%
  \parbox[t]{\@tabfnwidth}{\footnotesize #2\hfill}}
\catcode`\@=12 % @ signs are no longer letters
\newcommand{\dsp}        {\mbox{$D^{\ast +}$}}

\newcommand{\dssp}       {\mbox{$D_s^+$}}

\newcommand{\dz}         {\mbox{$D^{0}$}}
\newcommand{\dc}         {\mbox{$D^+$}}

\newcommand{\lc}         {\mbox{$\Lambda_c$}}
\newcommand{\lcp}        {\mbox{$\Lambda_c^+$}}

\newcommand{\fcdz}       {\mbox{$f(c \rightarrow D^0)$}}
\newcommand{\fcdc}       {\mbox{$f(c \rightarrow D^+)$}}
\newcommand{\fcdss}      {\mbox{$f(c \rightarrow D_s^+)$}}
\newcommand{\fclc}       {\mbox{$f(c \rightarrow \Lambda_c^+)$}}
\newcommand{\fcds}       {\mbox{$f(c \rightarrow D^{\ast +})$}}
\newcommand{\br}         {\mbox{${\cal B}_{D^{\ast +}\rightarrow D^0 \pi^+}$}}
\def\dsk3pi{ {\dsp}~\rightarrow~\dz~\pi^{+}_{s}%
        \rightarrow~(K^{-}~\pi^{+}~\pi^{+}~\pi^{-})~\pi^{+}_{s} }
\def\et10t{ E_T^{\theta > 10^\circ}}
\def\etw10{ E_T^{\theta > 10}}
\newcommand{\MeV}       {\mbox{${\,\rm MeV}$}}
\newcommand{\lum}{\mbox{$\cal L$}}
\newcommand{\bran}{\mbox{$\cal B$}}
\newcommand{\acc}{\mbox{$\cal A$}}
\def\citeCTD{{\cite{%
nim:a279:290,*npps:b32:181,*nim:a338:254%
}}\xspace}
\def\citeMVD{{\cite{%
nim:a581:656%
}}\xspace}
\def\citeSTT{{\cite{%
nim:a535:191%
}}\xspace}
\def\citeCAL{{\cite{%
nim:a309:77,*nim:a309:101,*nim:a321:356,*nim:a336:23%
}}\xspace}
\def\citeHES{{\cite{%
nim:a277:176%
}}\xspace}
\def\citeSixmT{{\cite{%
thesis:gosau:2007%
}}\xspace}
\def\citeBAC{{\cite{%
nim:a300:480%
}}\xspace}
\def\citePCAL{{\cite{%
desy-92-066,*zfp:c63:391,*acpp:b32:2025%
}}\xspace}
\def\citeSPECTRO{{\cite{%
nim:a565:572%
}}\xspace}
\begin{document}
%------------------------------------------------------------------------------
%       Title sheet
%------------------------------------------------------------------------------
%
% These are needed when draft mode is changed to preprint mode
%
\prepnum{DESY--13--106}
\prepdate{June 2013}

\date{}

\zeustitle{%
Measurement of charm fragmentation fractions in photoproduction at HERA
}
                    
\zeusauthor{ZEUS Collaboration}
\draftversion{2.3}
% \zeusdate{27 May 2011}

\maketitle

%
% If you use the package hepunits instead of units you have to enclose
% the values in curly brackets and change \pbi to \invpb and \Gev to \GeV
% e.g. \unit{47.7}{\invpb}
% If the quantity is not in math mode and the unit contains math mode
% characters such as superscripts it must be contained in $...$
%
\begin{abstract}\noindent
The production of $D^0$, $D^{*+}$, $D^{+}$, $D_s^{+}$ and $\Lambda_c^{+}$
charm hadrons and their antiparticles in
 $ep$ scattering at HERA has been studied with the ZEUS detector, using a 
total  integrated luminosity of 372 $\mbox{pb}^{-1}$. The fractions
  of charm quarks hadronising into a particular charm hadron were derived.
In addition, the ratio of neutral to charged 
$D$-meson production rates, the fraction of charged $D$ mesons produced in a 
vector state,
and the stangeness-suppression factor have been determined.
The measurements have been performed in the photoproduction regime. 
The charm hadrons were reconstructed in the range of
transverse momentum $p_T >3.8 \gev$ and pseudorapidity
$|\eta|<1.6$.
The charm fragmentation fractions are compared to previous results from HERA 
and
from $e^{+}e^{-}$ experiments. The data support the hypothesis that 
fragmentation is independent of the production process.
\end{abstract}
\thispagestyle{empty}
%------------------------------------------------------------------------------
%       Authors - you may have to play with \clearpage and \cleardoublepage 
%       in order to get the main text to start on the correct page
%------------------------------------------------------------------------------
\clearpage
%
%=================================================================== 
%
%  This file will be replaced with the "official" author
%  list which is created for each ZEUS paper individually.
%
%===================================================================== 
\pagenumbering{Roman} 

\begin{center} 
{\Large  The ZEUS Collaboration}
\end{center}

%===================================================================
%
%  MEMBER NAME  AUTH179 (ZEUS)     M  TEX
%
%  JH.: transformed to a format, which is suited as input for
%       CONVERT, which automatically creates author-indices
%
%  Don't remove lines starting with a percent sign %,
%  CONVERT may need them urgently !
%  
%=====================================================================
%
%\documentstyle[12pt,twoside]{report}  
%
%
%\topmargin-1.cm
%\evensidemargin-0.3cm
%\oddsidemargin-0.3cm
%\textwidth 16.cm
%\textheight 680pt
%\parindent0.cm
%\parskip0.3cm plus0.05cm minus0.05cm
%\def\3{\ss}
%\newcommand{\address}{ }
%\renewcommand{\author}{ }
%\pagenumbering{Roman}
%                                    % this "%"s are for cosmetics only
%\begin{document}
%                                                   %
%\begin{center}
%{                      \Large  The ZEUS Collaboration              }
%\end{center}

{\small\raggedright

%  members:

H.~Abramowicz$^{45, aj}$, 
I.~Abt$^{35}$, 
L.~Adamczyk$^{13}$, 
M.~Adamus$^{54}$, 
R.~Aggarwal$^{7, c}$, 
S.~Antonelli$^{4}$, 
P.~Antonioli$^{3}$, 
A.~Antonov$^{33}$, 
M.~Arneodo$^{50}$, 
O.~Arslan$^{5}$
% ...

\newpage

%       institutes:

\makebox[3em]{$^{1}$}
\begin{minipage}[t]{14cm}
{\it Argonne National Laboratory, Argonne, Illinois 60439-4815, USA}~$^{A}$

\end{minipage}\\
\makebox[3em]{$^{2}$}
\begin{minipage}[t]{14cm}
{\it Andrews University, Berrien Springs, Michigan 49104-0380, USA}

\end{minipage}\\
\makebox[3em]{$^{3}$}
\begin{minipage}[t]{14cm}
{\it INFN Bologna, Bologna, Italy}~$^{B}$

\end{minipage}\\
\makebox[3em]{$^{4}$}
\begin{minipage}[t]{14cm}
{\it University and INFN Bologna, Bologna, Italy}~$^{B}$

\end{minipage}\\
\makebox[3em]{$^{5}$}
\begin{minipage}[t]{14cm}
{\it Physikalisches Institut der Universit\"at Bonn,
Bonn, Germany}~$^{C}$

\end{minipage}\\
\makebox[3em]{$^{6}$}
\begin{minipage}[t]{14cm}
{\it H.H.~Wills Physics Laboratory, University of Bristol,
Bristol, United Kingdom}~$^{D}$

\end{minipage}\\
\makebox[3em]{$^{7}$}
\begin{minipage}[t]{14cm}
{\it Panjab University, Department of Physics, Chandigarh, India}

\end{minipage}\\
\makebox[3em]{$^{8}$}
\begin{minipage}[t]{14cm}
{\it Calabria University,
Physics Department and INFN, Cosenza, Italy}~$^{B}$

\end{minipage}\\
\makebox[3em]{$^{9}$}
\begin{minipage}[t]{14cm}
{\it Institute for Universe and Elementary Particles, Chonnam National University,\\
Kwangju, South Korea}

\end{minipage}\\
\makebox[3em]{$^{10}$}
\begin{minipage}[t]{14cm}
{\it Jabatan Fizik, Universiti Malaya, 50603 Kuala Lumpur, Malaysia}~$^{E}$

\end{minipage}\\
\makebox[3em]{$^{11}$}
\begin{minipage}[t]{14cm}
{\it Nevis Laboratories, Columbia University, Irvington on Hudson,
New York 10027, USA}~$^{F}$

\end{minipage}\\
\makebox[3em]{$^{12}$}
\begin{minipage}[t]{14cm}
{\it The Henryk Niewodniczanski Institute of Nuclear Physics, Polish Academy of \\
Sciences, Krakow, Poland}~$^{G}$

\end{minipage}\\
\makebox[3em]{$^{13}$}
\begin{minipage}[t]{14cm}
{\it AGH-University of Science and Technology, Faculty of Physics and Applied Computer
Science, Krakow, Poland}~$^{H}$

\end{minipage}\\
\makebox[3em]{$^{14}$}
\begin{minipage}[t]{14cm}
{\it Department of Physics, Jagellonian University, Cracow, Poland}

\end{minipage}\\
\makebox[3em]{$^{15}$}
\begin{minipage}[t]{14cm}
{\it Deutsches Elektronen-Synchrotron DESY, Hamburg, Germany}

\end{minipage}\\
\makebox[3em]{$^{16}$}
\begin{minipage}[t]{14cm}
{\it Deutsches Elektronen-Synchrotron DESY, Zeuthen, Germany}

\end{minipage}\\
\makebox[3em]{$^{17}$}
\begin{minipage}[t]{14cm}
{\it INFN Florence, Florence, Italy}~$^{B}$

\end{minipage}\\
\makebox[3em]{$^{18}$}
\begin{minipage}[t]{14cm}
{\it University and INFN Florence, Florence, Italy}~$^{B}$

\end{minipage}\\
\makebox[3em]{$^{19}$}
\begin{minipage}[t]{14cm}
{\it Fakult\"at f\"ur Physik der Universit\"at Freiburg i.Br.,
Freiburg i.Br., Germany}

\end{minipage}\\
\makebox[3em]{$^{20}$}
\begin{minipage}[t]{14cm}
{\it School of Physics and Astronomy, University of Glasgow,
Glasgow, United Kingdom}~$^{D}$

\end{minipage}\\
\makebox[3em]{$^{21}$}
\begin{minipage}[t]{14cm}
{\it Department of Engineering in Management and Finance, Univ. of
the Aegean, Chios, Greece}

\end{minipage}\\
\makebox[3em]{$^{22}$}
\begin{minipage}[t]{14cm}
{\it Hamburg University, Institute of Experimental Physics, Hamburg,
Germany}~$^{I}$

\end{minipage}\\
\makebox[3em]{$^{23}$}
\begin{minipage}[t]{14cm}
{\it Imperial College London, High Energy Nuclear Physics Group,
London, United Kingdom}~$^{D}$

\end{minipage}\\
\makebox[3em]{$^{24}$}
\begin{minipage}[t]{14cm}
{\it Institute of Particle and Nuclear Studies, KEK,
Tsukuba, Japan}~$^{J}$

\end{minipage}\\
\makebox[3em]{$^{25}$}
\begin{minipage}[t]{14cm}
{\it Institute of Physics and Technology of Ministry of Education and
Science of Kazakhstan, Almaty, Kazakhstan}

\end{minipage}\\
\makebox[3em]{$^{26}$}
\begin{minipage}[t]{14cm}
{\it Institute for Nuclear Research, National Academy of Sciences, Kyiv, Ukraine}

\end{minipage}\\
\makebox[3em]{$^{27}$}
\begin{minipage}[t]{14cm}
{\it Department of Nuclear Physics, National Taras Shevchenko University of Kyiv, Kyiv, Ukraine}

\end{minipage}\\
\makebox[3em]{$^{28}$}
\begin{minipage}[t]{14cm}
{\it Kyungpook National University, Center for High Energy Physics, Daegu,
South Korea}~$^{K}$

\end{minipage}\\
\makebox[3em]{$^{29}$}
\begin{minipage}[t]{14cm}
{\it Institut de Physique Nucl\'{e}aire, Universit\'{e} Catholique de Louvain, Louvain-la-Neuve,\\
Belgium}~$^{L}$

\end{minipage}\\
\makebox[3em]{$^{30}$}
\begin{minipage}[t]{14cm}
{\it Departamento de F\'{\i}sica Te\'orica, Universidad Aut\'onoma
de Madrid, Madrid, Spain}~$^{M}$

\end{minipage}\\
\makebox[3em]{$^{31}$}
\begin{minipage}[t]{14cm}
{\it Department of Physics, McGill University,
Montr\'eal, Qu\'ebec, Canada H3A 2T8}~$^{N}$

\end{minipage}\\
\makebox[3em]{$^{32}$}
\begin{minipage}[t]{14cm}
{\it Meiji Gakuin University, Faculty of General Education,
Yokohama, Japan}~$^{J}$

\end{minipage}\\
\makebox[3em]{$^{33}$}
\begin{minipage}[t]{14cm}
{\it Moscow Engineering Physics Institute, Moscow, Russia}~$^{O}$

\end{minipage}\\
\makebox[3em]{$^{34}$}
\begin{minipage}[t]{14cm}
{\it Lomonosov Moscow State University, Skobeltsyn Institute of Nuclear Physics,
Moscow, Russia}~$^{P}$

\end{minipage}\\
\makebox[3em]{$^{35}$}
\begin{minipage}[t]{14cm}
{\it Max-Planck-Institut f\"ur Physik, M\"unchen, Germany}

\end{minipage}\\
\makebox[3em]{$^{36}$}
\begin{minipage}[t]{14cm}
{\it NIKHEF and University of Amsterdam, Amsterdam, Netherlands}~$^{Q}$

\end{minipage}\\
\makebox[3em]{$^{37}$}
\begin{minipage}[t]{14cm}
{\it Physics Department, Ohio State University,
Columbus, Ohio 43210, USA}~$^{A}$

\end{minipage}\\
\makebox[3em]{$^{38}$}
\begin{minipage}[t]{14cm}
{\it Department of Physics, University of Oxford,
Oxford, United Kingdom}~$^{D}$

\end{minipage}\\
\makebox[3em]{$^{39}$}
\begin{minipage}[t]{14cm}
{\it INFN Padova, Padova, Italy}~$^{B}$

\end{minipage}\\
\makebox[3em]{$^{40}$}
\begin{minipage}[t]{14cm}
{\it Dipartimento di Fisica dell' Universit\`a and INFN,
Padova, Italy}~$^{B}$

\end{minipage}\\
\makebox[3em]{$^{41}$}
\begin{minipage}[t]{14cm}
{\it Department of Physics, Pennsylvania State University, University Park,\\
Pennsylvania 16802, USA}~$^{F}$

\end{minipage}\\
\makebox[3em]{$^{42}$}
\begin{minipage}[t]{14cm}
{\it Polytechnic University, Tokyo, Japan}~$^{J}$

\end{minipage}\\
\makebox[3em]{$^{43}$}
\begin{minipage}[t]{14cm}
{\it Dipartimento di Fisica, Universit\`a 'La Sapienza' and INFN,
Rome, Italy}~$^{B}$

\end{minipage}\\
\makebox[3em]{$^{44}$}
\begin{minipage}[t]{14cm}
{\it Rutherford Appleton Laboratory, Chilton, Didcot, Oxon,
United Kingdom}~$^{D}$

\end{minipage}\\
\makebox[3em]{$^{45}$}
\begin{minipage}[t]{14cm}
{\it Raymond and Beverly Sackler Faculty of Exact Sciences, School of Physics, \\
Tel Aviv University, Tel Aviv, Israel}~$^{R}$

\end{minipage}\\
\makebox[3em]{$^{46}$}
\begin{minipage}[t]{14cm}
{\it Department of Physics, Tokyo Institute of Technology,
Tokyo, Japan}~$^{J}$

\end{minipage}\\
\makebox[3em]{$^{47}$}
\begin{minipage}[t]{14cm}
{\it Department of Physics, University of Tokyo,
Tokyo, Japan}~$^{J}$

\end{minipage}\\
\makebox[3em]{$^{48}$}
\begin{minipage}[t]{14cm}
{\it Tokyo Metropolitan University, Department of Physics,
Tokyo, Japan}~$^{J}$

\end{minipage}\\
\makebox[3em]{$^{49}$}
\begin{minipage}[t]{14cm}
{\it Universit\`a di Torino and INFN, Torino, Italy}~$^{B}$

\end{minipage}\\
\makebox[3em]{$^{50}$}
\begin{minipage}[t]{14cm}
{\it Universit\`a del Piemonte Orientale, Novara, and INFN, Torino,
Italy}~$^{B}$

\end{minipage}\\
\makebox[3em]{$^{51}$}
\begin{minipage}[t]{14cm}
{\it Department of Physics, University of Toronto, Toronto, Ontario,
Canada M5S 1A7}~$^{N}$

\end{minipage}\\
\makebox[3em]{$^{52}$}
\begin{minipage}[t]{14cm}
{\it Physics and Astronomy Department, University College London,
London, United Kingdom}~$^{D}$

\end{minipage}\\
\makebox[3em]{$^{53}$}
\begin{minipage}[t]{14cm}
{\it Faculty of Physics, University of Warsaw, Warsaw, Poland}

\end{minipage}\\
\makebox[3em]{$^{54}$}
\begin{minipage}[t]{14cm}
{\it National Centre for Nuclear Research, Warsaw, Poland}

\end{minipage}\\
\makebox[3em]{$^{55}$}
\begin{minipage}[t]{14cm}
{\it Department of Particle Physics and Astrophysics, Weizmann
Institute, Rehovot, Israel}

\end{minipage}\\
\makebox[3em]{$^{56}$}
\begin{minipage}[t]{14cm}
{\it Department of Physics, University of Wisconsin, Madison,
Wisconsin 53706, USA}~$^{A}$

\end{minipage}\\
\makebox[3em]{$^{57}$}
\begin{minipage}[t]{14cm}
{\it Department of Physics, York University, Ontario, Canada M3J 1P3}~$^{N}$

\end{minipage}\\
\vspace{30em} \pagebreak[4]

%  references concerning institutes;

\makebox[3ex]{$^{ A}$}
\begin{minipage}[t]{14cm}
 supported by the US Department of Energy\
\end{minipage}\\
\makebox[3ex]{$^{ B}$}
\begin{minipage}[t]{14cm}
 supported by the Italian National Institute for Nuclear Physics (INFN) \
\end{minipage}\\
\makebox[3ex]{$^{ C}$}
\begin{minipage}[t]{14cm}
 supported by the German Federal Ministry for Education and Research (BMBF), under
 contract No. 05 H09PDF\
\end{minipage}\\
\makebox[3ex]{$^{ D}$}
\begin{minipage}[t]{14cm}
 supported by the Science and Technology Facilities Council, UK\
\end{minipage}\\
\makebox[3ex]{$^{ E}$}
\begin{minipage}[t]{14cm}
 supported by HIR and UMRG grants from Universiti Malaya, and an ERGS grant from the
 Malaysian Ministry for Higher Education\
\end{minipage}\\
\makebox[3ex]{$^{ F}$}
\begin{minipage}[t]{14cm}
 supported by the US National Science Foundation. Any opinion, findings and conclusions or
 recommendations expressed in this material are those of the authors and do not necessarily
 reflect the views of the National Science Foundation.\
\end{minipage}\\
\makebox[3ex]{$^{ G}$}
\begin{minipage}[t]{14cm}
 supported by the Polish Ministry of Science and Higher Education as a scientific project No.
 DPN/N188/DESY/2009\
\end{minipage}\\
\makebox[3ex]{$^{ H}$}
\begin{minipage}[t]{14cm}
 supported by the National Science Centre under contract No. DEC-2012/06/M/ST2/00428\
\end{minipage}\\
\makebox[3ex]{$^{ I}$}
\begin{minipage}[t]{14cm}
 supported by the German Federal Ministry for Education and Research (BMBF), under
 contract No. 05h09GUF, and the SFB 676 of the Deutsche Forschungsgemeinschaft (DFG) \
\end{minipage}\\
\makebox[3ex]{$^{ J}$}
\begin{minipage}[t]{14cm}
 supported by the Japanese Ministry of Education, Culture, Sports, Science and Technology
 (MEXT) and its grants for Scientific Research\
\end{minipage}\\
\makebox[3ex]{$^{ K}$}
\begin{minipage}[t]{14cm}
 supported by the Korean Ministry of Education and Korea Science and Engineering Foundation\
\end{minipage}\\
\makebox[3ex]{$^{ L}$}
\begin{minipage}[t]{14cm}
 supported by FNRS and its associated funds (IISN and FRIA) and by an Inter-University
 Attraction Poles Programme subsidised by the Belgian Federal Science Policy Office\
\end{minipage}\\
\makebox[3ex]{$^{ M}$}
\begin{minipage}[t]{14cm}
 supported by the Spanish Ministry of Education and Science through funds provided by CICYT\
\end{minipage}\\
\makebox[3ex]{$^{ N}$}
\begin{minipage}[t]{14cm}
 supported by the Natural Sciences and Engineering Research Council of Canada (NSERC) \
\end{minipage}\\
\makebox[3ex]{$^{ O}$}
\begin{minipage}[t]{14cm}
 partially supported by the German Federal Ministry for Education and Research (BMBF)\
\end{minipage}\\
\makebox[3ex]{$^{ P}$}
\begin{minipage}[t]{14cm}
 supported by RF Presidential grant N 3920.2012.2 for the Leading Scientific Schools and by
 the Russian Ministry of Education and Science through its grant for Scientific Research on
 High Energy Physics\
\end{minipage}\\
\makebox[3ex]{$^{ Q}$}
\begin{minipage}[t]{14cm}
 supported by the Netherlands Foundation for Research on Matter (FOM)\
\end{minipage}\\
\makebox[3ex]{$^{ R}$}
\begin{minipage}[t]{14cm}
 supported by the Israel Science Foundation\
\end{minipage}\\
\vspace{30em} \pagebreak[4]

%  references concerning mebers;

\makebox[3ex]{$^{ a}$}
\begin{minipage}[t]{14cm}
now at University of Salerno, Italy\
\end{minipage}\\
\makebox[3ex]{$^{ b}$}
\begin{minipage}[t]{14cm}
now at Queen Mary University of London, United Kingdom\
\end{minipage}\\
\makebox[3ex]{$^{ c}$}
\begin{minipage}[t]{14cm}
also funded by Max Planck Institute for Physics, Munich, Germany\
\end{minipage}\\
\makebox[3ex]{$^{ d}$}
\begin{minipage}[t]{14cm}
also Senior Alexander von Humboldt Research Fellow at Hamburg University,
 Institute of Experimental Physics, Hamburg, Germany\
\end{minipage}\\
\makebox[3ex]{$^{ e}$}
\begin{minipage}[t]{14cm}
also at Cracow University of Technology, Faculty of Physics,
 Mathematics and Applied Computer Science, Poland\
\end{minipage}\\
\makebox[3ex]{$^{ f}$}
\begin{minipage}[t]{14cm}
supported by the research grant No. 1 P03B 04529 (2005-2008)\
\end{minipage}\\
\makebox[3ex]{$^{ g}$}
\begin{minipage}[t]{14cm}
partially supported by the Polish National Science Centre projects DEC-2011/01/B/ST2/03643
 and DEC-2011/03/B/ST2/00220\
\end{minipage}\\
\makebox[3ex]{$^{ h}$}
\begin{minipage}[t]{14cm}
now at Rockefeller University, New York, NY
 10065, USA\
\end{minipage}\\
\makebox[3ex]{$^{ i}$}
\begin{minipage}[t]{14cm}
now at DESY group FS-CFEL-1\
\end{minipage}\\
\makebox[3ex]{$^{ j}$}
\begin{minipage}[t]{14cm}
now at Institute of High Energy Physics, Beijing, China\
\end{minipage}\\
\makebox[3ex]{$^{ k}$}
\begin{minipage}[t]{14cm}
now at DESY group FEB, Hamburg, Germany\
\end{minipage}\\
\makebox[3ex]{$^{ l}$}
\begin{minipage}[t]{14cm}
also at Moscow State University, Russia\
\end{minipage}\\
\makebox[3ex]{$^{ m}$}
\begin{minipage}[t]{14cm}
now at University of Liverpool, United Kingdom\
\end{minipage}\\
\makebox[3ex]{$^{ n}$}
\begin{minipage}[t]{14cm}
now at CERN, Geneva, Switzerland\
\end{minipage}\\
\makebox[3ex]{$^{ o}$}
\begin{minipage}[t]{14cm}
also affiliated with University College London, UK\
\end{minipage}\\
\makebox[3ex]{$^{ p}$}
\begin{minipage}[t]{14cm}
now at Goldman Sachs, London, UK\
\end{minipage}\\
\makebox[3ex]{$^{ q}$}
\begin{minipage}[t]{14cm}
also at Institute of Theoretical and Experimental Physics, Moscow, Russia\
\end{minipage}\\
\makebox[3ex]{$^{ r}$}
\begin{minipage}[t]{14cm}
also at FPACS, AGH-UST, Cracow, Poland\
\end{minipage}\\
\makebox[3ex]{$^{ s}$}
\begin{minipage}[t]{14cm}
partially supported by Warsaw University, Poland\
\end{minipage}\\
\makebox[3ex]{$^{ t}$}
\begin{minipage}[t]{14cm}
supported by the Alexander von Humboldt Foundation\
\end{minipage}\\
\makebox[3ex]{$^{ u}$}
\begin{minipage}[t]{14cm}
now at Istituto Nazionale di Fisica Nucleare (INFN), Pisa, Italy\
\end{minipage}\\
\makebox[3ex]{$^{ v}$}
\begin{minipage}[t]{14cm}
now at Haase Energie Technik AG, Neum\"unster, Germany\
\end{minipage}\\
\makebox[3ex]{$^{ w}$}
\begin{minipage}[t]{14cm}
now at Department of Physics, University of Bonn, Germany\
\end{minipage}\\
\makebox[3ex]{$^{ x}$}
\begin{minipage}[t]{14cm}
also affiliated with DESY, Germany\
\end{minipage}\\
\makebox[3ex]{$^{ y}$}
\begin{minipage}[t]{14cm}
also at University of Tokyo, Japan\
\end{minipage}\\
\makebox[3ex]{$^{ z}$}
\begin{minipage}[t]{14cm}
now at Kobe University, Japan\
\end{minipage}\\
\makebox[3ex]{$^{\dagger}$}
\begin{minipage}[t]{14cm}
 deceased \
\end{minipage}\\
\makebox[3ex]{$^{aa}$}
\begin{minipage}[t]{14cm}
supported by DESY, Germany\
\end{minipage}\\
\makebox[3ex]{$^{ab}$}
\begin{minipage}[t]{14cm}
member of National Technical University of Ukraine, Kyiv Polytechnic Institute,
 Kyiv, Ukraine\
\end{minipage}\\
\makebox[3ex]{$^{ac}$}
\begin{minipage}[t]{14cm}
member of National Technical University of Ukraine, Kyiv, Ukraine\
\end{minipage}\\
\makebox[3ex]{$^{ad}$}
\begin{minipage}[t]{14cm}
now at DESY ATLAS group\
\end{minipage}\\
\makebox[3ex]{$^{ae}$}
\begin{minipage}[t]{14cm}
member of National University of Kyiv - Mohyla Academy, Kyiv, Ukraine\
\end{minipage}\\
\makebox[3ex]{$^{af}$}
\begin{minipage}[t]{14cm}
Alexander von Humboldt Professor; also at DESY and University of Oxford\
\end{minipage}\\
\makebox[3ex]{$^{ag}$}
\begin{minipage}[t]{14cm}
STFC Advanced Fellow\
\end{minipage}\\
\makebox[3ex]{$^{ah}$}
\begin{minipage}[t]{14cm}
now at LNF, Frascati, Italy\
\end{minipage}\\
\makebox[3ex]{$^{ai}$}
\begin{minipage}[t]{14cm}
This material was based on work supported by the
 National Science Foundation, while working at the Foundation.\
\end{minipage}\\
\makebox[3ex]{$^{aj}$}
\begin{minipage}[t]{14cm}
also at Max Planck Institute for Physics, Munich, Germany, External Scientific Member\
\end{minipage}\\
\makebox[3ex]{$^{ak}$}
\begin{minipage}[t]{14cm}
now at Tokyo Metropolitan University, Japan\
\end{minipage}\\
\makebox[3ex]{$^{al}$}
\begin{minipage}[t]{14cm}
now at Nihon Institute of Medical Science, Japan\
\end{minipage}\\
\makebox[3ex]{$^{am}$}
\begin{minipage}[t]{14cm}
now at Osaka University, Osaka, Japan\
\end{minipage}\\
\makebox[3ex]{$^{an}$}
\begin{minipage}[t]{14cm}
also at \L\'{o}d\'{z} University, Poland\
\end{minipage}\\
\makebox[3ex]{$^{ao}$}
\begin{minipage}[t]{14cm}
member of \L\'{o}d\'{z} University, Poland\
\end{minipage}\\
\makebox[3ex]{$^{ap}$}
\begin{minipage}[t]{14cm}
now at Department of Physics, Stockholm University, Stockholm, Sweden\
\end{minipage}\\
}

%\end{document}
%
%
%
%
%{\raggedright
%  F.~Author$^{1}$,
%  A.~Nother,
%  X.~Whoever\\
% {\it Absolutely Non-Existing Institute of Nothing, Nowhere, Country}~$^{a}$ 
%}
%
%\par\filbreak
%\begin{supertabular}[h]{rp{14cm}}
%$^{\ 1}$ & now visiting a scientist next door\\ 
%\end{supertabular}
%
%\begin{tabular}[h]{rp{14cm}}
%$^{a}$ &  supported by the National Council of Nonsense (NCNS)\\ 
%\end{tabular} 

\clearpage
\pagenumbering{arabic}
%
% Comment out this line to remove date/time for final version
% 
%\pagestyle{scrheadings}

%
%==============================================================================
%       Here comes the document.
%==============================================================================
% ----------------------------------------------------------------------------
%       Introduction
% ----------------------------------------------------------------------------
\section{Introduction}
\label{sec-int}

 The fragmentation fractions of charm quarks into specific charm hadrons 
cannot be predicted by Quantum Chromodynamics (QCD) and have to be 
measured.  
It is usually assumed that they are universal, i.e. the same for 
charm quarks produced
in $e^{+}e^{-}$ annihilation, in $ep$ collisions and also in $pp$ 
or other hadronic collisions, even though the charm production
mechanisms are not the same: 
in $e^{+}e^{-}$ collisions, $c\bar{c}$ pairs are produced dominantly by QED
pair production, whereas in $ep$ collisions, the main production mechanism is 
the QCD boson-gluon fusion process $\gamma g \rightarrow c\bar{c}$.
The fragmentation universality can be tested by measuring the fragmentation 
fractions at HERA and comparing the results with those obtained with
$e^{+}e^{-}$ collisions.
Additionally, the values of the fragmentation fractions are crucial parameters
used in comparisons of perturbative QCD (pQCD) calculations with measurements
 of charm production at HERA and elsewhere.

In this paper, measurements of the photoproduction
of charm hadrons in $ep$ collisions at HERA are presented. 
The relative 
production rates of the most copiously produced charm ground states,
the $D^0$, $D^{+}$, $D_s^{+}$ mesons and the $\Lambda_c$ baryon, and of the $D^{*+}$ meson 
were measured\footnote{For all studied charm hadrons, the charge conjugated states are implied throughout the paper.}. 
The fractions of charm quarks hadronising into a particular
charm hadron, 
$f(c\rightarrow D, D^{*}, \Lambda_c)$ were determined in the kinematic 
range of transverse momentum $p_{T}(D, D^{*}, \Lambda_c) > 3.8 \gev$ and pseudorapidity
$|\eta(D, D^{*}, \Lambda_c)| < 1.6$ of the charm state.
Here $D$ stands for $D^0$, $D^{+}$ and $D_s^{+}$ mesons.
In addition, the ratio of neutral to charged
$D$-meson production rates, the fraction of charged $D$ mesons produced in a
vector state, and the strangeness-suppression factor were determined.

The analysis presented here is based on an independent data set with an integrated 
luminosity over 4.5 times larger than the previous ZEUS 
measurement~\cite{Chekanov:2005mm}.
The new measurement benefits also from the ZEUS microvertex detector (MVD), 
which made it possible to 
identify the secondary decay vertices
of the charm ground states and 
thereby to suppress background significantly. 
The new results are compared to the previous ZEUS measurement~\cite{Chekanov:2005mm}
in photoproduction, other HERA results from H1~\cite{Aktas:2004ka} and 
ZEUS~\cite{Chekanov:2007mm, Abramowicz:2010mm} in deep inelastic scattering,
and to results from experiments at $e^+e^-$ 
storage rings~\cite{Gladilin:Prep,Atlas:at2011}.
%
% ----------------------------------------------------------------------------
%       Experimental set-up
% ----------------------------------------------------------------------------
\section{Experimental set-up}
\label{sec-exp}

The analysis was performed with data taken from 2004 to 2007,
when HERA collided electrons or positrons with 
energy $E_e = 27.5\;$GeV and protons with energy
$E_p=920\;$GeV. The corresponding total integrated luminosity
was $372\pm 7\;\mbox{pb}^{-1}$.

%
% ----------------------------------------------------------------------------
%       General detector blabla
% ----------------------------------------------------------------------------
\Zdetdesc

% ----------------------------------------------------------------------------
%       CTD+MVD description, footnote on coordinate system is the argument
% ----------------------------------------------------------------------------
\Zctdmvddesc{\ZcoosysfnBEeta}

% ----------------------------------------------------------------------------
%       CAL description straight and simple
% ----------------------------------------------------------------------------
\Zcaldesc

% ----------------------------------------------------------------------------
%       Muon system description. You may want it or not. No references for it
% ----------------------------------------------------------------------------
%\Zmuondesc
%
% ----------------------------------------------------------------------------
%       BAC description complete with reference
% ----------------------------------------------------------------------------
%\Zbacdesc

% ----------------------------------------------------------------------------
%       LUMI, give actual luminosity uncertainty for your sample as
%       argument
%       Use \Zlumidesc is you don't want the uncertainty here.
% ----------------------------------------------------------------------------
\ZlumidescA{1.9}

%\vfill\eject

\section{Monte Carlo simulation}
\label{sec-simul}

Monte Carlo (MC) simulations were used in the analysis for
modelling signal and background processes and to correct
the data for acceptance effects.
MC samples of charm and beauty photoproduction events
were produced with
the {\sc Pythia} 6.416 event generator~\cite{Sjostrand:2006za}.
The generation of events, based on leading-order matrix elements,
includes direct photon processes,
in which the photon couples
as a point-like object in the hard scatter,
and resolved photon processes, where the photon acts as a source
of partons, one of which participates in the hard scattering process.
Initial- and final-state parton showering is added to simulate
higher-order processes.
The CTEQ5L~\cite{Lai:1999wy} and GRV~LO~\cite{Gluck:1991jc} parametrisations
were used for the parton distribution functions of the proton and photon, respectively.
The charm (beauty) quark masses were set to $1.5 \, (4.75)\;$GeV.
Events for all processes were generated in proportion to the predicted
MC cross sections.
The Lund string model~\cite{Andersson:1983ia}
as implemented in {\sc Jetset}~\cite{Sjostrand:2006za}
was used for hadronisation in {\sc Pythia}.
The Bowler modification~\cite{Bowler:1981sb}
of the Lund symmetric fragmentation function~\cite{Andersson:1983jt}
was used for the longitudinal component of the charm- and beauty-quark fragmentation.
The generated events were passed through a full simulation
of the detector using {\sc Geant} 3.21~\cite{Brun:1987ma}
and processed with the same reconstruction program as used for the data.

To ensure a good description of the data, 
a reweighting was applied to 
the transverse momentum, $p_T(D, D^{*}, \Lambda_c)$, and pseudorapidity,
$\eta(D, D^{*}, \Lambda_c)$, distributions of the
{\sc Pythia} MC samples. 
The reweighting
factors were tuned using a large $D^{*+}$ sample.
The factors deviate by no more than $\pm 15\%$ from unity.
The effect of the reweighting on the measured fragmentation
fractions was small; the reweighting uncertainty was included
in the systematic uncertainty. 
%
% ----------------------------------------------------------------------------
%      Section 4   Event Selection
% ----------------------------------------------------------------------------
\section{Event selection}
\label{sec-selec}

A three-level trigger system ~\cite{Smith:1992im} was used
to select events online.
The first- and second-level trigger used CAL and CTD
data to select $ep$ collisions and to reject beam-gas events.
At the third level, the full event information was available.
The sample used in this analysis was mainly selected by third-level triggers 
where at least one reconstructed charm-hadron candidate was required. A
dijet trigger was used in addition to increase the efficiency.

Photoproduction events were selected by requiring that no scattered
electron 
with energy of greater than 5 GeV be 
identified in the CAL~\cite{Derrick:1993tb}.
The photon-proton centre-of-mass energy, $W$, was reconstructed using the Jacquet-Blondel~\cite{Amaldi:1979yh} estimator of $W$,
${W_{\rm JB} = \sqrt{2 E_{p} \sum\limits_{i} E_{i}(1-\cos\theta_{i})}}$.
Here $E_{i}$ and $\theta_{i}$ denote the energy and polar angle
of the $i^{\rm th}$ energy-flow object (EFO)~\cite{Birskin:1998un}, respectively,
and the sum $i$ runs over all final-state energy-flow
objects built from CTD-MVD tracks and energy clusters
 measured in the CAL.
After correcting for detector effects, the most important of which were
energy losses in inactive material in front of the CAL and particle interactions in the beam   pipe~\cite{Derrick:1993tb, Derrick:1995sc}, 
events were selected in the interval $130<W_{\rm JB}<300\,$ GeV.
The lower limit was set by the trigger requirements, while the upper
limit was imposed to suppress remaining DIS events
with an unidentified low-energy scattered
electron in the CAL~\cite{Derrick:1993tb}.

\section{Reconstruction of charm hadrons}
\label{sec-rec}

The production yields of
$\dz$, $\dsp$, $\dc$, $\dssp$ and $\lcp$
charm hadrons
were measured in
the range of transverse momentum $p_T(D, D^{*}, \Lambda_c)>3.8\gev$
and the range of pseudorapidity
$|\eta(D, D^{*}, \Lambda_c)|<1.6$. 
%%%%%%%%%%
The $p_T$ cut was imposed by trigger requirements and the $\eta$ cut 
ensured a good acceptance in the CTD-MVD detector system.
%%%%%%%%%%
Charm hadrons were reconstructed using CTD-MVD tracks.
%%%%%%%%%%
Combinations of good tracks were used to form charm-hadron candidates,
as detailed in the following sections.
%%%%%%%%%% 
To ensure good momentum resolution, each track was required
to reach at least the third superlayer of the CTD.
The combinatorial background was significantly reduced by requiring
$p_T(D, D^{*})/\et10t > 0.2$
and $p_T(\Lambda_c)/\et10t > 0.25$
for charm mesons and baryons, respectively.
The transverse energy was calculated as
$\et10t={\Sigma_{i,\theta_i > 10^\circ}(E_{i}\sin \theta_i})$,
where the sum runs over all energy deposits in the CAL
with polar angles
$\theta_i$ above $10^\circ$. 
A further background reduction was achieved by applying cuts
on the minimal transverse momenta 
%and decay angles
of the charm-hadron decay products.
The large combinatorial background for the $\dz$, $\dc$ and $\dssp$
 mesons  
was additionally suppressed by secondary-decay vertex cuts
(see Section 5.1).

\subsection{Reconstruction of  $\pmb{D^{0}}$ mesons}
\label{sec-recd0}

The $\dz$ mesons were reconstructed
using the decay mode
$D^0 \rightarrow K^- \pi^+$.
In each event, tracks with opposite charges and
$p_T>0.8\gev$
were combined in pairs to form $\dz$ candidates.
The nominal kaon and pion masses were assumed
in turn for each track and
the invariant mass of the pair, $M(K \pi)$, was calculated.

The kaon and pion tracks, measured precisely in the CTD-MVD detector system,
were used to reconstruct the decay point of the $\dz$ meson. 
The relatively long lifetime of the $\dz$ meson resulted in a secondary vertex 
that is often well separated from the primary interaction point. 
This property was exploited to improve the signal-to-background ratio. 
The decay-length significance, $S_l$, was used as a discriminating variable.
It is defined as $S_{l} = l/\sigma_{l}$, where
$l$ is the decay length in the transverse plane and $\sigma_{l}$ is the uncertainty associated with this distance. 
The decay length is the distance
in the transverse plane between the point of creation and decay vertex of the meson and is given by

\begin{equation}
l = \frac{\left (\vec S_{XY} - \vec{B}_{XY} \right) \cdot \vec p^{D}_{T}}{p_{T}^{D}},
\end{equation}

where $\vec p^{D}_{T}$ is the transverse momentum vector and $\vec {S}_{XY}$
 is the two-dimensional 
position vector of the reconstructed decay vertex projected onto the $XY$ plane. 
The vector $\vec{B}_{XY}$ points to the fitted geometrical centre of the beam-spot 
which is taken as the origin of the $D$ meson. 
The centre of the elliptical beam-spot was determined 
using the average primary-vertex position for groups of a few thousand events. 
The vector $\vec{B}_{XY}$ was corrected for each event for the small difference in
angle between the beam direction and the $Z$ direction, using the $Z$ position
of the primary vertex of the event.
The widths of the beam spot were $88\mum$ ($80\mum$) and $24\mum$ ($22\mum$) in the $X$ and $Y$ directions, 
respectively, for the $e^{+}p$ ($e^{-}p$) data.
The decay-length error, $\sigma_{l}$, was
determined by folding the width of the beam-spot with the covariance matrix of the
decay vertex after both were projected onto the $D$-meson momentum vector.

A cut $S_l > 1$ was applied.
In addition, the $\chi^2$ of the vertex fit
was required to be less than 15; this quality cut was applied for
all secondary $D$-meson decay-vertex fits in this paper. 

For the selected $\dz$ candidates, a search was performed for a track
that could be a ``soft'' pion, $\pi_s$, from a $\dsp \rightarrow \dz \pi^+_s$
decay.
The soft pion was required to have $p_T>0.2\gev$ and a charge opposite
to that of the particle taken as a kaon.
The corresponding $\dz$ candidate was assigned to the class of candidates
``with $\Delta M$ tag''
if the mass difference,
$\Delta M=M(K \pi \pi_s)-M(K \pi)$, was in the range
$0.143<\Delta M<0.148\gev$.
All remaining $\dz$ candidates were assigned
to the class of candidates ``without $\Delta M$ tag''.

For $\dz$ candidates with $\Delta M$ tag,
the kaon and pion mass assignment was fixed
according to the charge of the tracks.
For $\dz$ candidates without $\Delta M$ tag, two mass
assignments were assumed for each $K \pi$ pair, yielding two entries into the
mass distribution: the true value, corresponding to the signal, and a wrong 
value, distributed over a broad range.
To remove this background,
the mass distribution, obtained for $\dz$ candidates with $\Delta M$ tag
and assigning the wrong masses to the kaon and pion tracks,
was subtracted from the $M(K \pi)$ distribution for all $\dz$ candidates
without $\Delta M$ tag. The subtracted mass distribution was normalised
to the ratio of numbers of $\dz$ mesons without and with
$\Delta M$ tag obtained from the fit described below.
Reflections from $D^0 \rightarrow K^- K^+$ and $D^0 \rightarrow \pi^- \pi^+$
 decays were seen as two small bumps below and above the signal peak,
 respectively, of the
 $D^0 \rightarrow K^- \pi^+$ decay. They were subtracted using the simulated
 reflection shapes and normalised 
to the $D^0 \rightarrow K^{-} \pi^{+}$ signal according to the normalisation ratios observed in the simulation and using the PDG values of
the respective branching ratios~\cite{Beringer:2012zzi}.

Figure 1 shows the $M(K \pi)$ distribution for $\dz$ candidates
with and without $\Delta M$ tag obtained after the subtractions described above.
Clear signals are seen at the nominal value of the $\dz$ mass in both
distributions.
The distributions were fitted simultaneously, assuming the same shape for
the signals in both distributions. To describe the shape, a modified
Gaussian function was used:
\begin{equation}
{\rm Gauss}^{\rm mod}\propto \exp [-0.5 \cdot x^{1+1/(1+0.5 \cdot x)}],
\label{eq:gausmod}
\end{equation}
where $x=|[M(K\pi)-M_0]/\sigma|$.
This functional form described both data and MC signals well.
The signal position, $M_0$,
and width, $\sigma$, and the number of $\dz$ mesons in each signal
were free parameters of the fit.
The background shape
in both distributions is compatible with being
approximately linear in the mass range above $1.92\gev$.
For smaller $M(K \pi)$ values, there is an enhancement due to contributions 
from other $\dz$ decay modes and other $D$ mesons, as was verified 
by the Monte Carlo simulation.

The background shape in the fit was described by the form
$[A+B\cdot M(K \pi)]$ for $M(K \pi)>1.92\gev$ and
$[A+B\cdot M(K \pi)]\cdot \exp\{D\cdot{[M(K \pi)-1.92]}^2\}$
for $M(K \pi)<1.92\gev$. 
The free parameters $A$, $B$ and $D$ were assumed to be independent for the
two $M(K \pi)$ distributions.
The numbers of $\dz$ mesons yielded by the fit were $N^{\rm tag}(\dz)=7281\pm104$ and
$N^{\rm untag}(\dz)=27787\pm680$ for selections with and without $\Delta M$ tag,
respectively. The mass value obtained from the fit\footnote{For all fitted mass values 
in this paper the quoted uncertainties are only statistical.}
 was $1865.4\pm0.3$ MeV for the $\dz$ tagged and $1865.1\pm0.4$ MeV for the $\dz$ untagged
samples, compared to the PDG value
of $1864.83 \pm 0.14$ MeV~\cite{Beringer:2012zzi}.

\subsection{Reconstruction of additional $\pmb{D^{*+}}$ mesons}
\label{sec-recds}

The $\dsp \rightarrow \dz \pi^+_s$  decays
with $p_T(\dsp)>3.8\gev$ and $|\eta(\dsp)|<1.6$
can be considered as a sum of two subsamples:
decays with the $\dz$ having $p_T(\dz)>3.8\gev$ and $|\eta(\dz)|<1.6$,
and decays with the $\dz$ outside that kinematic range.
The former sample is represented by
$\dz$ mesons reconstructed with $\Delta M$ tag, as discussed in the previous
section. The latter sample of additional $\dsp$ mesons
was obtained using
the same $D^0 \rightarrow K^- \pi^+$ decay channel
and the selection described below.

In each event, tracks with opposite charges and $p_T>0.4\gev$
were combined in pairs to form $\dz$ candidates.
To calculate the invariant mass, $M(K \pi)$,
kaon and pion masses were assumed in turn for each track.
Only $\dz$ candidates which satisfy $1.81<M(K \pi)<1.92\gev$
were kept.
Moreover, the $\dz$ candidates were required to have
$p_T(\dz)<3.8\gev$ or $|\eta(\dz)|>1.6$.
Any additional track with $p_T>0.2\gev$
and a charge opposite to that of the kaon track
was assigned the pion mass and combined with the $\dz$ candidate
to form a $\dsp$ candidate
with invariant mass $M(K \pi \pi_s)$.

Figure 2 shows
the $\Delta M = M(K\pi\pi_s) - M(K\pi)$ distribution
for the
$\dsp$ candidates from the additional $D^{*}$-meson subsample after all cuts.
A clear signal is seen at the nominal value of
$M(\dsp)-M(\dz)$. 
The sum of the modified
Gaussian function
(Eq.~(\ref{eq:gausmod}))
describing the signal
and a function of the form $A\cdot (\Delta M -m_\pi)^B\cdot 
e^{-C\cdot\Delta M}$, describing the non-resonant background, was used to fit 
the data. Here $m_\pi$ is the pion mass
and $A$, $B$ and $C$ are free parameters of the fit.
The fitted mass value$^{3}$ for the $\Delta M$ signal is ${145.51 \pm 0.01\MeV}$, compared
to the PDG value of ${145.42 \pm 0.01\MeV}$~\cite{Beringer:2012zzi}. 
The number of reconstructed additional $D^{*+}$ mesons determined from the fit
was ${N^{\rm add}(\dsp)=2139\pm59}$.

The combinatorial background was estimated also from
the mass-difference distribution for wrong-charge combinations,
in which both tracks forming the $\dz$ candidate had the same charge
and the third track had the opposite charge.
The number of reconstructed additional $D^{*+}$ mesons was determined by 
subtracting the wrong-charge $\Delta M$ distribution after normalising it
to the distribution of $D^{*+}$ candidates with the appropriate charges
in the range $0.151 < \Delta M < 0.167$ GeV. The subtraction was performed in the signal range $0.143 < \Delta M < 0.148$ GeV.
The results obtained using the subtraction
procedure instead of the fit were used to estimate the systematic uncertainty
of the signal extraction.

\subsection{Reconstruction of $\pmb{D^{+}}$ mesons}
\label{sec-recdc}

The $\dc$ mesons were reconstructed
using the decay mode
$D^+ \rightarrow K^-\pi^+\pi^+$.
In each event, two tracks with the same charge and
$p_T>0.5\gev$ and
a third track with the opposite charge
and $p_T>0.7\gev$
were combined to form $\dc$ candidates.
The pion mass was assigned to the two tracks
with the same charge, the kaon mass was assigned to the third track,
and the candidate invariant mass, $M(K\pi\pi)$, was calculated.
To suppress background from $\dsp$ decays, combinations with
$M(K\pi\pi)-M(K\pi)<0.15\gev$ were removed.
The background from
$\dssp \rightarrow \phi\pi^+$ with $\phi \rightarrow K^+K^-$
was suppressed by requiring that the invariant mass of any two
tracks with opposite charges from $\dc$ candidates
was not within $\pm8\, \mev$ of the $\phi$ mass~\cite{Beringer:2012zzi}
when the kaon mass was assigned to both tracks.
To suppress combinatorial background, a cut on the decay-length significance 
for $\dc$ candidates was applied of $S_l > 3$.

Figure 3 shows the $M(K\pi\pi)$ distribution for
the $\dc$ candidates after all cuts.
A clear signal is seen at the nominal value of the $D^+$ mass.
The sum of two Gaussian functions with the same peak position
was used to describe the signal:
\begin{equation}
{\rm Gauss}^{\rm sum} = \frac{p_{0}}{\sqrt{2\pi}}  
\left[\, p_{3}/p_{2} \cdot \exp [-(x - p_{1})^2/2p_{2}^{2}] + (1 - p_{3})/p_{4}
 \cdot \exp [ -(x - p_{1})^2/2p_{4}^{2} ]\;\right],
\label{eq:gaussum}
\end{equation}
where $x=M(K\pi\pi)$.

An exponential function describing the non-resonant background was used.
Reflections caused by wrong mass assignments 
for the decay products of $\dssp$ and $\lcp$
decaying to three charged particles
were added to the fit function using the simulated reflection shapes 
normalised to the measured $\dssp$ and $\lcp$ production rates.
They give rise to a small increase of the background in the signal region.
The number of reconstructed $\dc$ mesons yielded by the fit was
$N(\dc)=18917\pm324$. The fitted mass$^{3}$ 
of the  $\dc$
was $1869.0\pm0.2\mev$, compared to
the PDG value of $1869.62 \pm 0.15\mev$~\cite{Beringer:2012zzi}.

\subsection{Reconstruction of $\pmb{D^{+}_s}$ mesons}
\label{sec-recdss}

The $\dssp$ mesons were reconstructed
using the decay mode
$\dssp \rightarrow \phi\pi^+$ with $\phi \rightarrow K^+K^-$.
In each event, tracks with opposite charges and
$p_T>0.7\gev$
were assigned the kaon mass and
combined in pairs to form $\phi$ candidates.
The $\phi$ candidate was kept if its invariant mass, $M(KK)$,
was within $\pm8 \,\mev$ of the $\phi$ mass~\cite{Beringer:2012zzi}.
Any additional track  with $p_T>0.5\gev$
was assigned the pion mass and combined with the $\phi$ candidate
to form a $\dssp$ candidate
with invariant mass $M(K K \pi)$. The cut on the decay-length significance for
 $\dssp$ candidates was $S_l > 0$.

Figure 4 shows the $M(K K\pi)$ distribution for
the $\dssp$ candidates after all cuts.
A clear signal is seen at the nominal $\dssp$ mass.
There is also a smaller signal around the nominal $\dc$ mass
as expected from the decay
$\dc \rightarrow \phi\pi^+$ with $\phi \rightarrow K^+K^-$.
The mass distribution was fitted by the sum of two modified
Gaussian functions
(Eq.~(\ref{eq:gausmod}))
describing the signals
and an exponential function describing the non-resonant background.
To reduce the number of free fit parameters in the fit,
the ratio 
of the widths of the $D^+$ and  $\dssp$ signals
was fixed to the value observed in the MC simulation.
Reflections arising from wrong mass assignments for the decay products of $\dc$ and $\lcp$
decays to three
charged particles
were added to the fit function using the simulated reflection shapes
normalised to the measured $\dc$ and $\lcp$ production rates.
The number of reconstructed $\dssp$
mesons yielded by the fit was
$N(\dssp)=2802\pm141$.
The fitted mass$^{3}$ 
of the $\dssp$  
was $1968.0\pm0.5$ MeV, compared to 
the PDG value of $1968.49 \pm 0.32$ MeV~\cite{Beringer:2012zzi}.

\subsection{Reconstruction of $\pmb{\lcp}$ baryons}
\label{sec-reclc}

The $\lcp$ baryons were reconstructed
using the decay mode
$\lcp \rightarrow K^-p\pi^+$.
In each event, two same-charge tracks
and a third track with opposite charge
were combined to form $\lcp$ candidates.
Due to the large difference between the proton and pion masses
and the high $\lcp$ momentum, the proton
momentum
is typically larger than that of
the pion.
Therefore, the proton (pion) mass was assigned to the track of
the same-charge pair with the larger (smaller) momentum. 
The kaon mass was assigned to the third track and
the invariant mass, $M(K p\pi)$, was calculated.
Only candidates with $p_T(K)>0.5\gev$, $p_T(p)>1.3\gev$
and $p_T(\pi)>0.5\gev$ were kept.
Reflections from $\dc$ and $\dssp$
decays to three
charged particles were subtracted from the $M(K p\pi)$ spectrum using the simulated reflection shapes
normalised to the measured $\dc$ and $\dssp$ production rates.

Figure 5 shows the $M(K p\pi)$ distribution for
the $\lcp$ candidates after all cuts, obtained after the reflection 
subtraction.
A clear signal is seen at the nominal $\lcp$ mass.
The sum of a modified
Gaussian function
(Eq.~(\ref{eq:gausmod}))
describing the signal
and a background function parametrised as 
\begin{equation*}
\exp[A \cdot M(Kp\pi) +B] \cdot M(Kp\pi)^C, 
\end{equation*}
where $\it{A, B}$ and $\it{C}$ are free parameters, was fitted
to the mass distribution.
The width parameter of the modified Gaussian
was fixed to $\sigma = 10\;$MeV.
This value corresponds to the width determined
in the MC, multiplied by a factor $1.11$.
The uncertainty of this number is taken into account in the systematics 
variations. 
The factor $1.11$ corrects for the difference
of the observed width of the $D^+ \rightarrow K^-\pi^+\pi^+$ signal between
data and simulation.
The number of reconstructed $\lcp$ baryons yielded by the fit was
$N(\lcp)=7682\pm964$.
The fitted mass$^{3}$ 
of the $\lcp$  
was $2290\pm1.8$ MeV, compared to
the PDG value of $2286.46 \pm 0.14$ MeV~\cite{Beringer:2012zzi}.

\section{Charm-hadron production cross sections}
\label{sec-xsec}

The cross sections for the production of the various charm hadrons
were determined, but results involve only ratios, in which
common normalisation uncertainties cancel. 

The fraction of charm quarks hadronising as a particular charm hadron,
$f(c\rightarrow D, D^{*}, \Lambda_c)$, is given by the ratio
of the production cross section for the hadron to the sum
of the production cross sections
for all charm ground states.
The charm-hadron cross sections were determined
for the process  $e p\rightarrow e(D, D^{*},\Lambda_c) X$
in the kinematic region $Q^2<1\gev^2$,
$\,\,\,130<W<300\gev$,
$p_T(D, D^{*},\Lambda_c)>3.8\gev$ and $|\eta(D, D^{*}, \Lambda_c)|<1.6$.

The cross section for a given charm hadron was calculated from
\begin{equation}
\sigma(D, D^{*}, \Lambda_c)=\frac{ N^{\rm{data}}_{(D, D^{*}, \Lambda_c)} -  s_b \cdot N^{b,\rm{MC}}_{(D, D^{*}, \Lambda_c)} }{\acc \cdot \lum \cdot \bran}\,,
\label{eq:xsecform}
\end{equation}
where $N^{\rm{data}}_{(D, D^{*}, \Lambda_c)}$ denotes the number of reconstructed charm hadrons in the data,
$\acc$ the acceptance for this charm hadron,
$\lum$ the integrated luminosity
and $\bran$ the branching ratio or the product of the
branching ratios~\cite{Beringer:2012zzi}
for the decay channels used in the reconstruction.
The {\sc Pythia} MC sample of charm photoproduction (see Section~\ref{sec-simul})
was used to evaluate the acceptance.
The contributions from beauty-hadron decays were subtracted
using the prediction from {\sc Pythia}.
For this purpose, 
the branching ratios of beauty-quark decays to the charmed hadrons
were corrected in the MC, using the correction factors ~\cite{Chekanov:2005mm}
based on the values measured at LEP~\cite{Buskulic:1996ah,Ackerstaff:1997ki}.
Finally, the number of reconstructed charm hadrons from beauty,
 $N^{b,{\rm MC}}_{D,D^*,\Lambda_c}$,
in the MC, normalised to the data luminosity and multiplied by 
a scale factor, $s_b$, was subtracted from the data (Eq.~\ref{eq:xsecform}).
The scale factor was chosen as $s_b=1.5 \pm 0.5$,
an average value which was estimated from ZEUS measurements~\cite{ZEUS:2011aa,Chekanov:2008tx,Chekanov:2008aaa}
of beauty photoproduction.

Using the number of reconstructed signal events (see Section~\ref{sec-rec}),
the following cross sections
for the sum of each charm hadron and its antiparticle
were calculated:

\begin{itemize}
\item{
for $\dz$ mesons
not originating from  $\dsp \rightarrow \dz \pi^+_s$ decays, 
$\sigma^{\rm untag}(\dz)$;
}
\item{
for $\dz$ mesons 
from $\dsp \rightarrow \dz \pi^+_s$ decays,
$\sigma^{\rm tag}(\dz)$.
The ratio $\sigma^{\rm tag}(\dz)/\br$ gives
the $\dsp$
cross section, $\sigma(\dsp)$,
corresponding to $\dz$ production in the kinematic range
$p_T(\dz)>3.8\gev$ and $|\eta(\dz)|<1.6$
for the $\dsp \rightarrow \dz \pi^+_s$ decay.
Here $\br=0.677$
is the branching ratio of the $\dsp \rightarrow \dz \pi^+_s$ decay~\cite{Beringer:2012zzi};
}
\item{
for additional $\dsp$ mesons,
$\sigma^{\rm add}(\dsp)$. 
The sum $\sigma^{\rm tag}(\dz)/\br + \sigma^{\rm add}(\dsp)$ gives
the $\dsp$ cross section, $\sigma^{\rm kin}(\dsp)$,
corresponding to $\dsp$ production in the kinematic range
$p_T(\dsp)>3.8\gev$ and $|\eta(\dsp)|<1.6$;
}
\item{
for $\dc$ mesons,
$\sigma(\dc)$;
}
\item{
for $\dssp$ mesons,
$\sigma(\dssp)$;
}
\item{
for $\lcp$ baryons,
$\sigma(\lcp) $.
}
\end{itemize}

\section{Systematic uncertainties}
\label{sec-syst}

The systematic uncertainties were determined by changing the analysis procedure
or by varying parameter values within their estimated uncertainties.
The following systematic uncertainty sources
were considered:
\begin{itemize}
\item{$\{\delta_1\}$
the uncertainty of the beauty subtraction (see Section~\ref{sec-xsec}) was determined by
varying the scale factor $s_b$ for the {\sc Pythia} MC prediction by $\pm 0.5$ 
from the nominal value $s_b=1.5$.
This was done to account for the range of the {\sc Pythia} beauty-prediction scale 
factors extracted in various analyses~\cite{ZEUS:2011aa,Chekanov:2008tx,Chekanov:2008aaa}.
In addition 
the branching ratios of
$b$ quarks to
charm hadrons were varied by their uncertainties~\cite{Buskulic:1996ah,Ackerstaff:1997ki};
}
\item{$\{\delta_2\}$
the uncertainty in the rate
of the charm-strange baryons (see Section~\ref{sec-ff}) was determined by varying the 
normalisation factor for the $\lcp$ production cross section 
by its estimated uncertainty~\cite{Chekanov:2005mm} of $\pm 0.05$
from the nominal value 1.14;
}
\item{$\{\delta_3\}$
the uncertainties related to the signal extraction procedures 
(see Sections 5.1--5.5) were
obtained by the following (independent) variations:}
\begin{itemize}
\item
for the $\dz$ signals with and without $\Delta M$ tag:
the background parametrisation was changed: for the region 
$M(K\pi)<1.92\;$GeV a linear term $C\cdot [M(K \pi)-1.92]$
was added to the argument of the exponential function; the transition point for the parametrisation 
was moved from $1.92\gev$ to $1.84\gev$.
The fit range was narrowed by $50\;$MeV on both sides;
\item
for the additional $\dsp$ signal:
the $M(K\pi)$ mass window for the selected $\dz$ candidates was narrowed by 
5.5 MeV on both sides.
The range used for the fit of the $\Delta M$ distribution 
was narrowed by 1 MeV (left) and 5 MeV (right);

The wrong-charge subtraction procedure was used instead of the fit; the range 
used for
the normalisation of the wrong-charge background was narrowed by
$1\;$MeV (left) and $5\;$MeV (right); the signal range used 
for the wrong-charge subtraction was narrowed or broadened by $1\;$MeV on both sides;
\item
for the $\dc$ signal:
a modified Gaussian was used as an alternative parametrisation
for the signal; the background parametrisation was changed
to a parabola. 
The fit range was narrowed by $50\;$MeV on both sides;
\item 
for the  $\dssp$ signal:
the background parametrisation was changed
to a parabola. 
The fit range was narrowed by $50\;$ MeV (left) and $30\;$ MeV (right);
\item 
for the $\lcp$ signal:
the background parametrisation
was changed to a cubic polynomial.
The fit range was narrowed by $30\;$ MeV on both sides.
The width parameter $\sigma$ of the modified Gaussian 
(Eq.~(\ref{eq:gausmod})) was varied by $\pm 10\%$ from its nominal value,
a conservative estimate of its uncertainty.
Further cross checks were performed: the width of the modified Gaussian was used as
a free fit parameter; the mass of the $\lcp$ was fixed to the PDG value~\cite{Beringer:2012zzi}.
The resulting signal-yield changes from these two variations
were negligible.
\end{itemize}
The uncertainties arising from the various reflections in the mass spectra
(see Section~\ref{sec-rec})
were evaluated by varying the size of each reflection
conservatively by $\pm20\%$.

The largest contribution to the signal extraction procedures was the change of
the background parametrisation;
\item{$\{\delta_4\}$
the model dependence of the acceptance corrections was estimated 
by varying the reweighting of the MC kinematic distributions (see Section~\ref{sec-simul})
until clear discrepancies became visible between the shapes observed in the 
data and in the MC;
}
\item{$\{\delta_5\}$
the uncertainty of the trigger efficiency was evaluated by comparing the fitted
signal yields taken with independent triggers. This uncertainty largely 
cancels in the fragmentation fractions;
}
\item{$\{\delta_6\}$
the track-finding inefficiency in the data with respect to the MC was estimated to be at most 2$\%$. This leads to a possible underestimation
of the production cross sections for the charm hadrons
with two (three) decay tracks by a factor $1.02^2$ ($1.02^3$)
which was taken into account for the systematics of the
fragmentation fractions; 
}
\item{$\{\delta_7\}$
the uncertainty of the CAL simulation was determined by
varying the simulation: the CAL energy scale was changed by $\pm 2\%$ and
the CAL energy resolution by $\pm 20\%$ of its value;
}
\item{$\{\delta_8\}$
the uncertainty related to the $S_l$ cut was 
determined by changing the value of the cut to $S_l>4$ for $\dc$ and
by omitting the $S_l$ cut for $\dz$ and $\dssp$.
}
\end{itemize}

Contributions from
the different systematic uncertainties were calculated and added
in quadrature separately for positive and negative variations.
The total and individual systematic uncertainties $\delta_1$ to $\delta_8$
for the charm fragmentation fractions
are summarised in
Table~\ref{tab:syst}.

The largest systematic uncertainties are related to the signal-extraction 
procedures.

\section{Results}
\label{sec-ratio}

\subsection{Equivalent phase-space treatment}
\label{sec-equ}

To compare the inclusive $D^+$ and $D^0$ cross sections with each other and
with the inclusive $D^{*+}$ cross section, it is necessary to take into account
 that in the $D^{*}$ decay 
only a fraction of the parent $D^*$ momentum is transferred to the daughter $D$ 
meson. For such a comparison, the ``equivalent''  $D^+$ and $D^0$ cross
sections, 
$\sigma^{\rm eq}(D^+)$ and $\sigma^{\rm eq}(D^0)$,
were defined [1] as the cross section
for $D^+$ and $D^0$ production including the contributions from $D^*$ decay, 
plus the contribution from additional $D^*$ mesons (see Section~5.2).
The cross section for $D^+$ and $D^0$ production is $\sigma(D^+)$ and 
$\sigma^{\rm tag}(D^0) + \sigma^{\rm untag}(D^0)$, respectively. 
The contributions from
additional $D^*$ mesons 
are, for the $D^+$ meson,
\begin{equation*}
\sigma^{\rm add}(D^+)
=
\sigma^{\rm add}(D^{*+})
\cdot(1-B_{D^{*+} \to D^0 \pi^+})
\end{equation*}
and for the $D^0$ meson
\begin{equation*}
\sigma^{\rm add}(D^0)=\sigma^{\rm add}(D^{*+})B_{D^{*+} \to D^{0} \pi^{+}}+
\sigma^{\rm add}(D^{*0}),
\end{equation*}
noting that $D^{*0}$ decays always to $D^0$~\cite{Beringer:2012zzi}.  

The cross-section $\sigma^{\rm add}(D^{*0})$ is not measured and is determined
as
\begin{equation}
\sigma^{\rm add}(D^{*0})=\sigma^{\rm add}(D^{*+})\cdot R_{u/d},
\label{eq:dsadd}
\end{equation}
where $R_{u/d}$ is the ratio of neutral to charged $D$-meson production rates.
It is given by the ratio of the sum of $D^{*0}$ and direct $D^0$ production
to the sum of $D^{*+}$ and direct $D^+$ production cross sections. 
It can be written as~\cite{Chekanov:2005mm} 
\begin{equation}
R_{u/d}=\frac{\sigma^{\rm untag}(D^0)}{\sigma(D^+)+\sigma^{\rm tag}(D^0)}.
\label{eq:rud}
\end{equation}
Combining everything produces the following expressions for 
$\sigma^{\rm eq}(D^+)$ and $\sigma^{\rm eq}(D^0)$:
\begin{equation*}
\sigma^{\rm eq}(D^+)=\sigma(D^+)+\sigma^{\rm add}(D^+)=\sigma(D^+)+
\sigma^{\rm add}(D^{*+})
\cdot(1-B_{D^{*+} \to D^0 \pi^+})
\end{equation*}
and
\begin{eqnarray*}
\sigma^{\rm eq}(D^0) & = & \sigma^{\rm untag}(D^0)+\sigma^{\rm tag}(D^0)+
\sigma^{\rm add}(D^0)\\
& = & \sigma^{\rm untag}(D^0)+\sigma^{\rm tag}(D^0)+\sigma^{\rm add}(D^{*+})
B_{D^{*+} \to D^{0} \pi^{+}}+
\sigma^{\rm add}(D^{*0}),
\end{eqnarray*}
which together with Eq.~\eqref{eq:dsadd} gives
\begin{equation*}
\sigma^{\rm eq}(D^0)=\sigma^{\rm untag}(D^0)+\sigma^{\rm tag}(D^0)+
\sigma^{\rm add}(D^{*+})
\cdot(
B_{D^{*+} \to D^{0} \pi^{+}}+R_{u/d}).
\end{equation*}
The observable $R_{u/d}$ was measured 
in the kinematic region $Q^2<1\gev^2$,
$130<W<300\gev$,
$p_T(D)>3.8\gev$ and $|\eta(D)|<1.6$.
The value 
obtained from Eq.~\eqref{eq:rud} is
$$R_{u/d}= 1.09 \pm 0.03 \,({\rm stat.}) ^{+0.04}_{-0.03}\,({\rm syst.})
\pm 0.02\,({\rm br}),$$
where the last uncertainty arises from the uncertainties of the branching ratios used.
The result is in agreement with the previous 
measurement~\cite{Chekanov:2005mm}
and slightly above but still compatible with $R_{u/d} = 1$,
expected from isospin invariance in the kinematic range of this measurement. 

Monte Carlo studies performed for the  previous ZEUS 
measurement~\cite{Chekanov:2005mm} 
showed that this equivalent phase-space treatment
for the non-strange  $D$ and $D^*$ mesons
minimises differences between the fragmentation
fractions measured in the accepted $p_T(D,D^*,\Lambda_c)$ and 
$\eta(D,D^*,\Lambda_c)$
kinematic region and those in the
full phase space. The extrapolation factors using the 
{\sc Pythia} MC with either the Peterson or Bowler fragmentation function
were generally close to unity to within a few percent~\cite{Chekanov:2005mm}.
 
\subsection{Charm fragmentation fractions}
\label{sec-ff}

For the determination of the fragmentation fractions of the
$\dz$, $\dc$, $\dssp$ and $\lcp$ charm ground states, 
the production cross sections of the charm-strange baryons
$\Xi^{+}_c$, $\Xi^{0}_c$ and $\Omega^0_c$
were taken into account.
The production rates for these baryons are expected
to be much lower than that of the $\lcp$
due to strangeness suppression.
The relative rates for the
ground states of the charm-strange baryons
were estimated from the non-charm sector
following the LEP procedure~\cite{Opal:97}.
The total rate for the three charm-strange baryons relative to the
$\lcp$ state is expected to be about $14\%$~\cite{Chekanov:2005mm}.
Therefore the $\lcp$ production cross section was scaled by the
factor $1.14$.

Using the equivalent $\dz$ and $\dc$ cross sections,
the sum
of the production cross sections
for all open-charm ground states, $\sigma_{\rm gs}$, is given by
$$\sigma_{\rm gs} = \sigma^{\rm eq}(D^+) + \sigma^{\rm eq}(D^0) + \sigma(D_s^+)
+ \sigma(\Lambda_c^+)\cdot 1.14,$$
which can be expressed using $R_{u/d}$ from Eq.~(\ref{eq:rud}) as
\begin{eqnarray*}
\sigma_{\rm gs} = \sigma(D^+) + \sigma^{\rm untag}(D^0) +
\sigma^{\rm tag}(D^0) +
\sigma^{\rm add}(D^{*+})\cdot (1+R_{u/d}) + \sigma(D_s^+) 
+ \sigma(\Lambda_c^+)\cdot 1.14. 
\end{eqnarray*}

The fragmentation fractions for the measured charm ground states and for 
$D^{*+}$ are given by
\begin{eqnarray*}
f(c\rightarrow D^+)&=&\sigma^{\rm eq}(D^+)/\sigma_{\rm gs}
=[\sigma(D^+)+\sigma^{\rm add}(D^{*+})\cdot (1-\br)]/\sigma_{\rm gs},\\
f(c\rightarrow D^0)&=&\sigma^{\rm eq}(D^0)/\sigma_{\rm gs}\\
&=&[\sigma^{\rm untag}(D^0)+\sigma^{\rm tag}(D^0)+\sigma^{\rm add}(D^{*+})\cdot (R_{u/d}+\br)]/\sigma_{\rm gs},\\
f(c\rightarrow D_s^+)&=&\sigma(D_s^+)/\sigma_{\rm gs},\\
f(c\rightarrow \Lambda_c^+)&=&\sigma(\Lambda_c^+)/\sigma_{\rm gs},\\
f(c\rightarrow D^{*+})&=&\sigma^{\rm kin}(\dsp)/\sigma_{\rm gs}=[\sigma^{\rm tag}(D^0)/\br+\sigma^{\rm add}(D^{*+})]/\sigma_{\rm gs}.
\end{eqnarray*}

The charm fragmentation fractions,
measured in the kinematic
region $Q^2<1\gev^2$,
$130<W<300\gev$,
$p_T(D, D^{*}, \Lambda_c)>3.8\gev$ and $|\eta(D, D^{*}, \Lambda_c)|<1.6$,
are summarised in
Table~\ref{tab:ff}. These results have been computed using the PDG 2012 
branching-ratio values~\cite{Beringer:2012zzi}.
The measurements
are compared to
previous HERA results~\cite{Chekanov:2005mm, Aktas:2004ka, Chekanov:2007mm, Abramowicz:2010mm} and
to the combined fragmentation fractions
for charm production in $e^+e^-$ annihilations compiled
previously~\cite{Gladilin:Prep}
and updated~\cite{Atlas:at2011, Lohrmann:2011fr} with the 2010 branching-ratio 
values~\cite{Nakamura:2010zzi}. This comparison is also shown in Fig. 6.
The obtained precision of the fragmentation fractions is competitive
with measurements in $e^+ e^-$ collisions.
All data from $ep$ and $e^{+}e^{-}$ collisions are in agreement with each 
other.
This demonstrates that the
fragmentation fractions of charm quarks are independent of the production
process and supports the hypothesis of universality of heavy-quark 
fragmentation.

The charm fragmentation fractions can also be
 used~\cite{Chekanov:2005mm} to determine
the fraction of charged $D$ mesons produced in a vector state, $P_v^d$, and
the strangeness-suppression factor, $\gamma_s$:
\begin{equation*}
P_v^d=\frac{\sigma^{\rm kin}(\dsp)}{\sigma^{\rm kin}(\dsp)+\sigma^{\rm dir}(\dc)} = 
\frac{\sigma^{\rm tag}(\dz)/\br+\sigma^{\rm add}(D^{*+})}{\sigma(D^{+})+
\sigma^{\rm tag}(\dz)+\sigma^{\rm add}(D^{*+})}
\end{equation*}
and
\begin{equation*}
\gamma_s=\frac{2\sigma(D^+_s)}{\sigma^{\rm eq}(D^+)+\sigma^{\rm eq}(D^0)}.
\end{equation*}
The value of $P_v^d$ obtained is
\begin{equation*}
P_v^d=0.595\pm0.020({\rm stat.})\pm0.015({\rm syst.})\pm 0.011({\rm br.}).
\end{equation*}
This is consistent with the result from the previous 
publication~\cite{Chekanov:2005mm} and with the result from combined $e^+e^-$
data~\cite{Gladilin:Prep,Atlas:at2011}. It is smaller than the naive
spin-counting prediction of 0.75 and also smaller than 2/3, the value predicted by the string-fragmentation
approach~\cite{Pei:1996kq}.

The strangeness-suppression factor obtained is 
\begin{equation*}
\gamma_s=0.214\pm0.013({\rm stat.})^{+0.006}_{-0.017}({\rm syst.})\pm 0.012
({\rm br.}),
\end{equation*}
consistent with the result from the previous 
publication~\cite{Chekanov:2005mm}. It is interesting to compare this value,
derived from charm decays, with values
derived from strange particle production, which are between 0.22 and 
0.3~\cite{Opal:1995,Aleph:1996,Ham:95,Chekanov:2007k,Soe:1988}.

\section{Summary}
\label{sec-conc}
The photoproduction of the charm hadrons
$\dz$, $\dsp$, $\dc$, $\dssp$ and $\lcp$ and their corresponding antiparticles
has been measured with the ZEUS detector
in in the kinematic range
$p_T(D, D^{*}, \Lambda_c)>3.8\gev$, $|\eta(D, D^{*}, \Lambda_c)|<1.6$,
$130<W<300\gev$ and $Q^2<1\gev^2$.

Using a data set with an integrated luminosity of 372 $\mbox{pb}^{-1}$,
the fractions of charm quarks hadronising as
$\dz$, $\dsp$, $\dc$, $\dssp$ and $\lcp$
hadrons have been determined. 
In addition, the ratio of neutral to charged $D$-meson production rates, the 
fraction of charged $D$ mesons produced in a vector state, and the 
strangeness-suppression factor have been determined.

The precision of the fragmentation fractions obtained is competitive
with measurements in $e^+ e^-$ collisions.
All data from $ep$ and $e^{+}e^{-}$ collisions are in agreement with each 
other.
This demonstrates that the
fragmentation fractions of charm quarks are independent of the production
process and supports the hypothesis of the universality of heavy-quark 
fragmentation.

% ----------------------------------------------------------------------------
%       Mandatory acknowledgements. You may add your buddies to it.
% ----------------------------------------------------------------------------
\section*{Acknowledgements}
\label{sec-ack}

\Zacknowledge
\vfill\eject

{
\ifzeusbst
  \bibliographystyle{./BiBTeX/bst/l4z_default}
\fi
\ifzdrftbst
  \bibliographystyle{./BiBTeX/bst/l4z_draft}
\fi
\ifzbstepj
  \bibliographystyle{./BiBTeX/user/l4z_epj-ICB}
\fi
\ifzbstnp
  \bibliographystyle{./BiBTeX/bst/l4z_np}
\fi
\ifzbstpl
  \bibliographystyle{./BiBTeX/bst/l4z_pl}
\fi
{\raggedright
\bibliography{./BiBTeX/user/syn.bib,%
              ./BiBTeX/user/myref.bib,%
              ./BiBTeX/bib/l4z_zeus.bib,%
              ./BiBTeX/bib/l4z_h1.bib,%
              ./BiBTeX/bib/l4z_articles.bib,%
              ./BiBTeX/bib/l4z_books.bib,%
              ./BiBTeX/bib/l4z_conferences.bib,%
              ./BiBTeX/bib/l4z_misc.bib,%
              ./BiBTeX/bib/l4z_preprints.bib}}
}
\vfill\eject

%-------------------------------------------------------------------------------
%       An example table
%-------------------------------------------------------------------------------
% \begin{table}[p]
% \begin{center}
% \begin{tabular}{||c|rrrr|rrrr||}
% \hline
% Label & $\eta^u_{LL}$ & $\eta^u_{LR}$ & $\eta^u_{RL}$ & $\eta^u_{RR}$ & 
%         $\eta^d_{LL}$ & $\eta^d_{LR}$ & $\eta^d_{RL}$ & $\eta^d_{RR}$ \\ 
% \hline\hline
% VV&  $+1$&$+1$&$+1$&$+1$&
%      $+1$&$+1$&$+1$&$+1$ \\
% AA&  $+1$&$-1$&$-1$&$+1$&
%      $+1$&$-1$&$-1$&$+1$ \\
% VA&  $+1$&$-1$&$+1$&$-1$&
%      $+1$&$-1$&$+1$&$-1$ \\
% X1&  $+1$&$-1$&$ 0$&$ 0$&
%      $+1$&$-1$&$ 0$&$ 0$ \\
% X2&  $+1$&$ 0$&$+1$&$ 0$&
%      $+1$&$ 0$&$+1$&$ 0$ \\
% X3&  $+1$&$ 0$&$ 0$&$+1$&
%      $+1$&$ 0$&$ 0$&$+1$ \\
% X4&  $ 0$&$+1$&$+1$&$ 0$&
%      $ 0$&$+1$&$+1$&$ 0$ \\
% X5&  $ 0$&$+1$&$ 0$&$+1$&
%      $ 0$&$+1$&$ 0$&$+1$ \\
% X6&  $ 0$&$ 0$&$+1$&$-1$&
%      $ 0$&$ 0$&$+1$&$-1$ \\
% U1&  $+1$&$-1$&$ 0$&$ 0$&
%      $ 0$&$ 0$&$ 0$&$ 0$ \\
% U2&  $+1$&$ 0$&$+1$&$ 0$&
%      $ 0$&$ 0$&$ 0$&$ 0$ \\
% U3&  $+1$&$ 0$&$ 0$&$+1$&
%      $ 0$&$ 0$&$ 0$&$ 0$ \\
% U4&  $ 0$&$+1$&$+1$&$ 0$&
%      $ 0$&$ 0$&$ 0$&$ 0$ \\
% U5&  $ 0$&$+1$&$ 0$&$+1$&
%      $ 0$&$ 0$&$ 0$&$ 0$ \\
% U6&  $ 0$&$ 0$&$+1$&$-1$&
%      $ 0$&$ 0$&$ 0$&$ 0$ \\
% \hline
% \end{tabular}
% \caption{The 30 scenarios for a better life considered in this paper.
%          Each row of this table corresponds to two different scenarios for
%          overall interference signs $\epsilon=+1$ and $\epsilon=-1$, 
%          respectively.}
%   \label{tab-models}
% \end{center}
% \end{table}

\begin{table}[hbt!]
\begin{center}
\begin{tabular}{|c|c|c|c|} \hline
& ZEUS ($\gamma p$)  & ZEUS ($\gamma p$) [1] & ZEUS (DIS) [3,4] \\
& HERA II & HERA I & HERA I\\
\hline
\hline
&\phantom{~~~~~~~~~} stat.\phantom{~~} syst.\phantom{~~~} br.
&\phantom{~~~~~~~~~} stat.\phantom{~~} syst.\phantom{~~~} br.
&\phantom{~~~~~~~~~} stat.\phantom{~~} syst.\phantom{~~} br.\\
\hline
% \rule{0pt}{1cm}
$\fcdc$ & $0.234 \pm 0.006\phantom{~}^{+0.004}_{-0.006}\phantom{~}^{+0.006}_{-0.008}$ & $0.222 \pm 0.015\phantom{~}^{+0.014}_{-0.005}\phantom{~}^{+0.011}_{-0.013}$  &  $0.217 \pm 0.018\phantom{~}^{+0.002}_{-0.019}\phantom{~}^{+0.009}_{-0.010}$\\
\hline
$\fcdz$& $0.588 \pm 0.017\phantom{~}^{+0.011}_{-0.006}\phantom{~}^{+0.012}_{-0.018}$ & $0.532 \pm 0.022\phantom{~}^{+0.018}_{-0.017}\phantom{~}^{+0.019}_{-0.028}$ & $0.585 \pm 0.019\phantom{~}^{+0.009}_{-0.052}\phantom{~}^{+0.018}_{-0.019}$ \\
\hline
$\fcdss$& $0.088 \pm 0.006\phantom{~}^{+0.002}_{-0.007}\phantom{~}^{+0.005}_{-0.005}$ & $0.075 \pm 0.007\phantom{~}^{+0.004}_{-0.004}\phantom{~}^{+0.005}_{-0.005}$ & $0.086 \pm 0.010\phantom{~}^{+0.007}_{-0.008}\phantom{~}^{+0.005}_{-0.005}$ \\
\hline
$\fclc$ & $0.079 \pm 0.013\phantom{~}^{+0.005}_{-0.009}\phantom{~}^{+0.024}_{-0.014}$ & $0.150 \pm 0.023\phantom{~}^{+0.014}_{-0.022}\phantom{~}^{+0.038}_{-0.025}$ & $0.098 \pm 0.027\phantom{~}^{+0.020}_{-0.017}\phantom{~}^{+0.025}_{-0.023}$\\
\hline
\hline
$\fcds$& $0.234 \pm 0.006\phantom{~}^{+0.004}_{-0.004}\phantom{~}^{+0.005}_{-0.007}$ & $0.203 \pm 0.009\phantom{~}^{+0.008}_{-0.006}\phantom{~}^{+0.007}_{-0.010}$ & $0.234 \pm 0.011\phantom{~}^{+0.006}_{-0.021}\phantom{~}^{+0.007}_{-0.010}$ \\
\hline
\end{tabular}
\end{center}
\end{table}

\begin{table}[hbt!]
\begin{center}
\begin{tabular}{|c|c|c|} \hline
& H1 (DIS) [2] & Combined \\
&  & $e^+e^-$ data [5,6]\\
\hline
\hline
&\phantom{~~~~}stat.$\oplus\,$syst.\phantom{~} br.
&\phantom{~~~~}stat.$\oplus\,$syst.\phantom{~} br.\\
\hline
$\fcdc$& $0.204 \pm 0.026\phantom{~}^{+0.009}_{-0.010}$ & $0.222\phantom{~} \pm 0.010\phantom{~}^{+0.010}_{-0.009}$\\
\hline
$\fcdz$& $0.584 \pm 0.048\phantom{~}^{+0.018}_{-0.019}$  & $0.544\phantom{~} \pm 0.022\phantom{~}^{+0.007}_{-0.007}$\\
\hline
$\fcdss$& $0.121 \pm 0.044\phantom{~}^{+0.008}_{-0.008}$ & $0.077\phantom{~} \pm 0.006\phantom{~}^{+0.005}_{-0.004}$\\
\hline
$\fclc$&  & $0.076\phantom{~} \pm 0.007\phantom{~}^{+0.027}_{-0.016}$\\
\hline
\hline
$\fcds$& $0.276 \pm 0.034\phantom{~}^{+0.009}_{-0.012}$ & $0.235\phantom{~} \pm 0.007\phantom{~}^{+0.003}_{-0.003}$\\
\hline
\end{tabular}
\caption{
Fractions of charm quarks hadronising as a particular charm hadron,
$f(c \rightarrow D, D^*, \Lambda_c)$.
The fractions are shown for the $D^+$, $D^0$, $D^+_s$
and $\Lambda_c^+$ charm ground states and for the
$D^{*+}$ state. The fractions in this and the previous ZEUS paper [1] were 
determined for the
kinematic range $p_T > 3.8$ GeV, $|\eta| < 1.6$ and $130<W<300\,$GeV.
Data for previous results [1-6] were updated to 2010 branching 
ratios [6,32,33]; data
from this paper were calculated with 2012 branching ratios [25].
}
\label{tab:ff}
\end{center}
\end{table}

\begin{table}[hbt]
\begin{center}
\begin{tabular}{|c|c|c|c|c|c|c|c|c|c|} \hline
& total & $\delta_1$
& $\delta_2$
& $\delta_3$
& $\delta_4$
& $\delta_5$
& $\delta_6$
& $\delta_7$
& $\delta_8$
\\
& $(\%)$ & $(\%)$ 
& $(\%)$
& $(\%)$
& $(\%)$
& $(\%)$
& $(\%)$
& $(\%)$
& $(\%)$
\\
\hline
\hline

$\fcdc$
&
$^{+1.8}_{-2.7}$
&
$^{+0.3}_{-0.3}$
&
$^{+0.4}_{-0.4}$
&
$^{+1.4}_{-2.0}$
&
$^{+0.3}_{-0.3}$
&
$^{+0.6}_{-0.6}$
&
${+1.0}$
&
$^{+0.2}_{-1.6}$
&
$^{+0.2}_{-0.1}$
\\
\hline
$\fcdz$
&
$^{+1.7}_{-1.0}$
&
$^{+0.2}_{-0.2}$
&
$^{+0.4}_{-0.4}$
&
$^{+1.6}_{-0.6}$
&
$^{+0.1}_{-0.1}$
&
$^{+0.3}_{-0.3}$
&
${-0.7}$
&
${+0.8}$
&
$^{+0.2}_{-0.1}$
\\
\hline
$\fcdss$
&
$^{+2.1}_{-8.0}$
&
$^{+0.4}_{-0.4}$
&
$^{+0.4}_{-0.3}$
&
$^{+1.3}_{-7.6}$
&
$^{+0.1}_{-0.1}$
&
$^{+0.8}_{-0.9}$
&
${+1.1}$
&
$^{+0.3}_{-1.9}$
&
$^{+0.2}_{-0.1}$
\\
\hline
$\fclc$
&
$^{+6.4}_{-11.7}$
&
$^{+0.1}_{-0.1}$
&
$^{+0.4}_{-0.3}$
&
$^{+6.1}_{-11.6}$
&
$^{+0.2}_{-0.1}$
&
$^{+1.1}_{-0.4}$
&
${+1.0}$
&
$^{+0.5}_{-0.9}$
&
${-0.7}$
\\
\hline
$\fcds$
&
$^{+1.9}_{-1.9}$
&
$^{+1.0}_{-1.0}$
&
$^{+0.4}_{-0.4}$
&
$^{+1.5}_{-1.6}$
&
$^{+0.2}_{-0.1}$
&
$^{+0.4}_{-0.4}$
&
${-0.4}$
&
$^{+0.3}_{-0.1}$
&
${+0.2}$
\\
\hline
\end{tabular}
\caption{
The total and individual $\delta_1$--$\delta_8$ (see text)
systematic uncertainties
for the charm-hadron
 fragmentation fractions.
}
\label{tab:syst}
\end{center}
\end{table}

\clearpage
%-------------------------------------------------------------------------------
%       Results
%-------------------------------------------------------------------------------
\begin{figure}[p]
\vfill
\begin{center}
\includegraphics[scale=0.80]{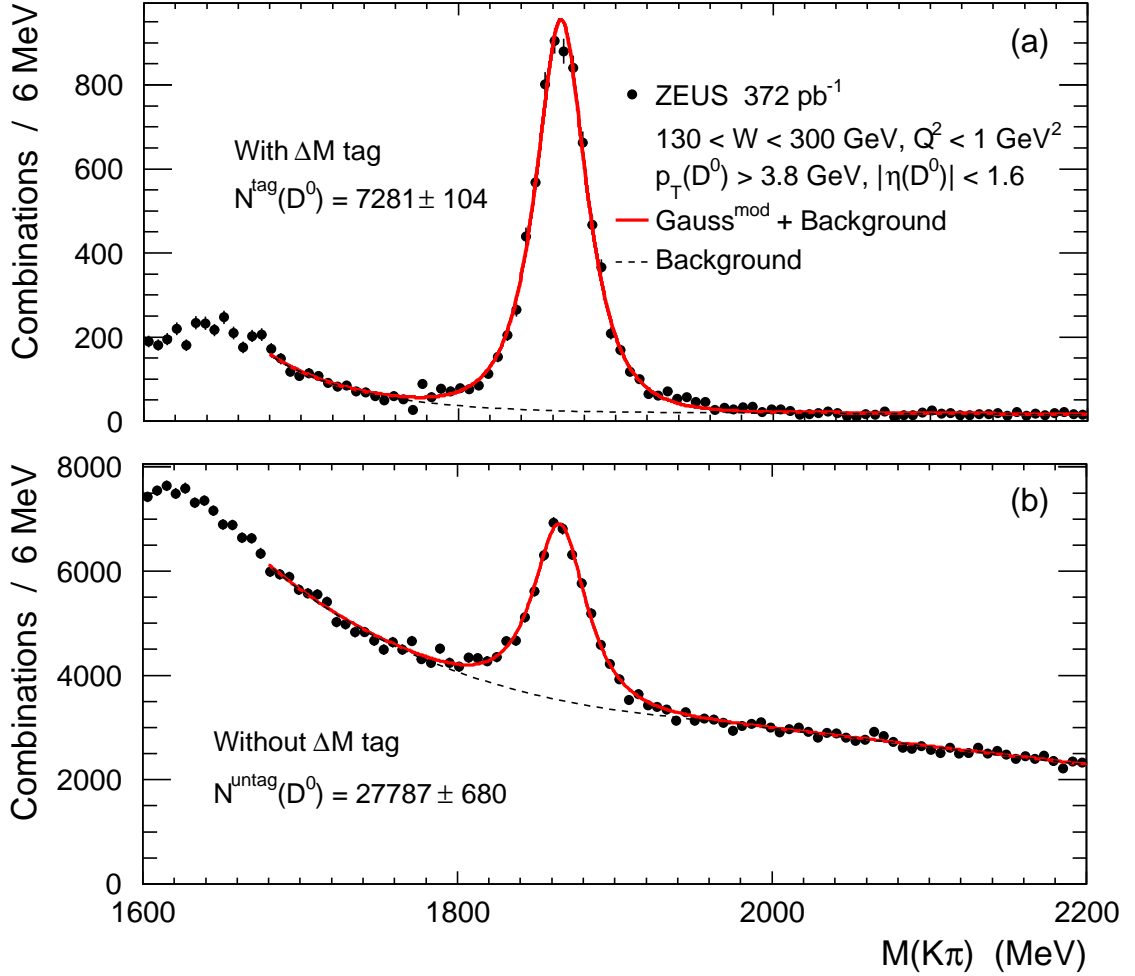}
\end{center}
\caption{The
$M(K\pi)$ distribution (dots) for (a) the $\dz$ candidates with $\Delta M$ tag,
and for (b) the $\dz$ candidates without $\Delta M$ tag,
 obtained after the subtractions described in the text.
The solid curves represent a fit to the sum of a modified Gaussian function 
and a background function (see text).
The background is also shown separately (dashed curves).
}
\label{1}
\end{figure}

\begin{figure}[p]
\vfill
\begin{center}
\includegraphics[scale=0.80]{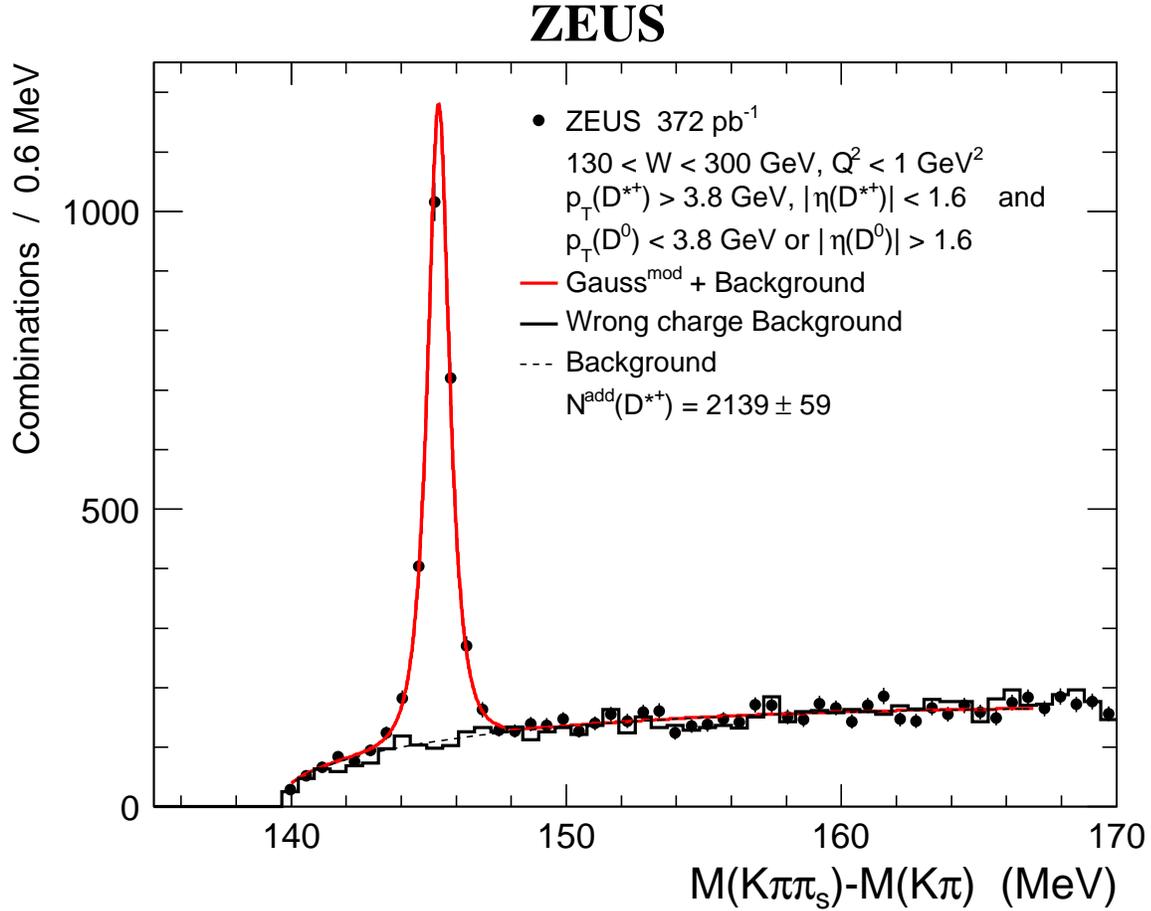}
\end{center}
\caption{
The distribution of the mass difference, $\Delta M = M(K\pi\pi_s) - M(K\pi)$,
 for the additional $\dsp$ candidates (dots).
The histogram solid shows the $\Delta M$ distribution for wrong-charge 
combinations.   
The solid curve represents a fit to the sum of a modified Gaussian function
and a background function (see text).
The background is also shown separately (dashed curve).
}
\label{1}
\end{figure}

\begin{figure}[p]
\vfill
\begin{center}
\includegraphics[scale=0.80]{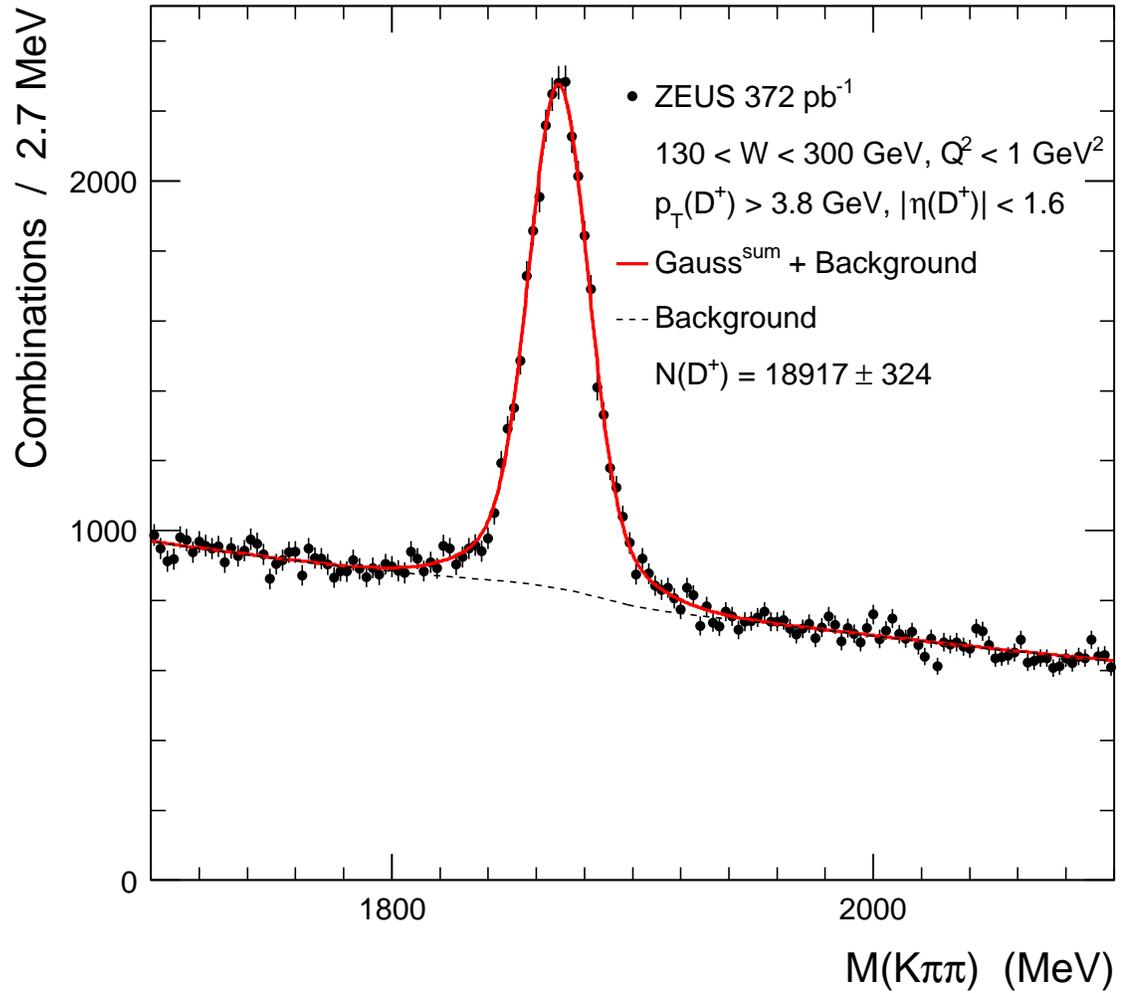}
\end{center}
\caption{The
$M(K\pi\pi)$ distribution for the $\dc$ candidates (dots).
The solid curve represents a fit to the sum of two Gaussian functions
and a background function. The background (dashed curve) is
a sum of an exponential function and reflections from decays of other
charm hadrons (see text). The reflections give rise to a small increase 
of the background in the signal region.}
\label{1}
\end{figure}

\begin{figure}[p]
\vfill
\begin{center}
\includegraphics[scale=0.80]{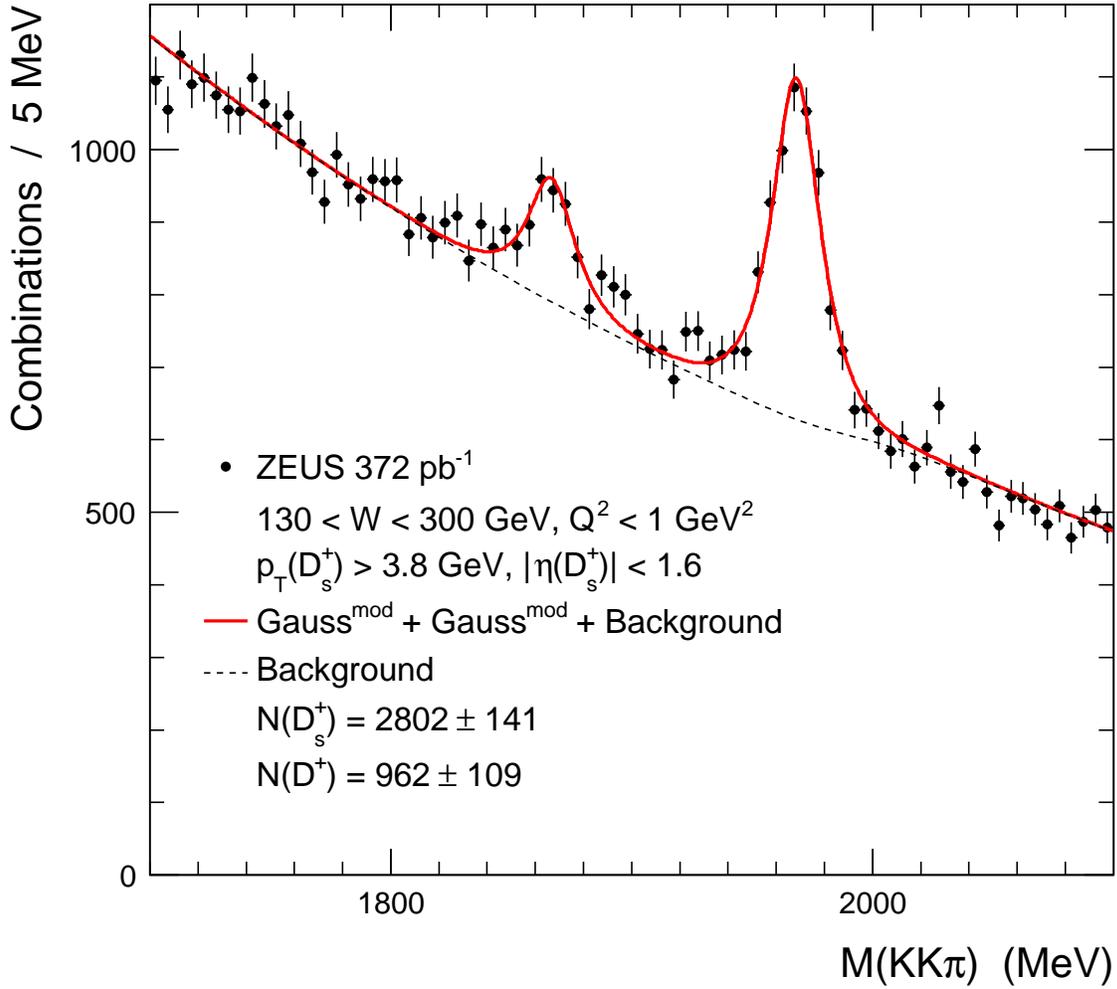}
\end{center}
\caption{The
$M(KK\pi)$ distribution for the $\dssp$ candidates (dots).
The solid curve represents a fit to the sum of two modified Gaussian functions 
and a background function. The peak at 1870  MeV is due to the 
decay $D^+ \to K^+ K^- \pi^+$.
The background (dashed curve) is
a sum of an exponential function and reflections from decays of other
charm hadrons (see text). 
}
\label{1}
\end{figure}

\begin{figure}[p]
\vfill
\begin{center}
\includegraphics[scale=0.80]{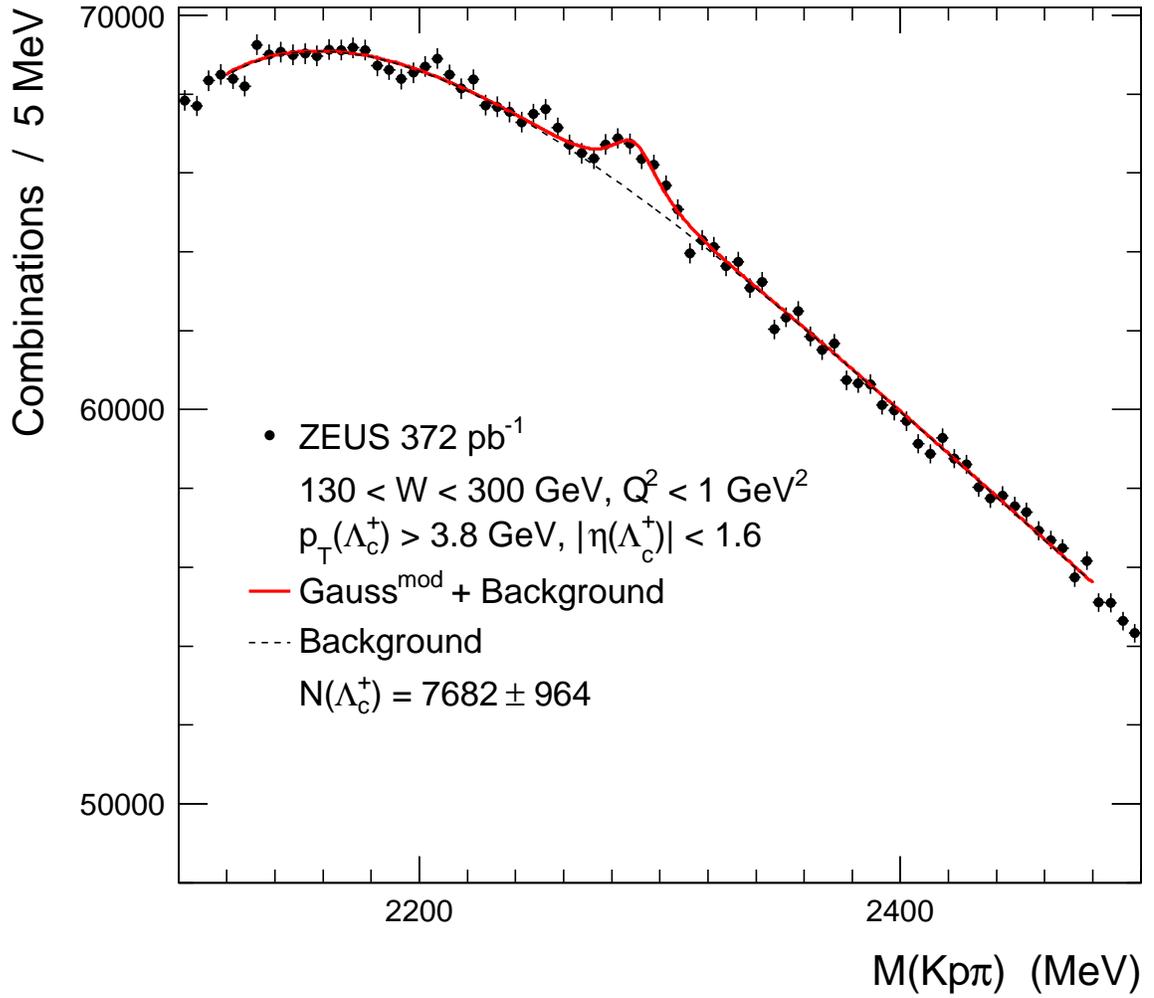}
\end{center}
\caption{The
$M(Kp\pi)$ distribution for the $\lc$ candidates (dots),
obtained after reflection subtraction (see text).
The solid curve represents a fit to the sum of a modified Gaussian function 
and a background function (see text). The background is also shown separately
 (dashed curve).
}
\label{1}
\end{figure}

\begin{figure}[p]
\vfill
\begin{center}
\includegraphics[scale=0.90]{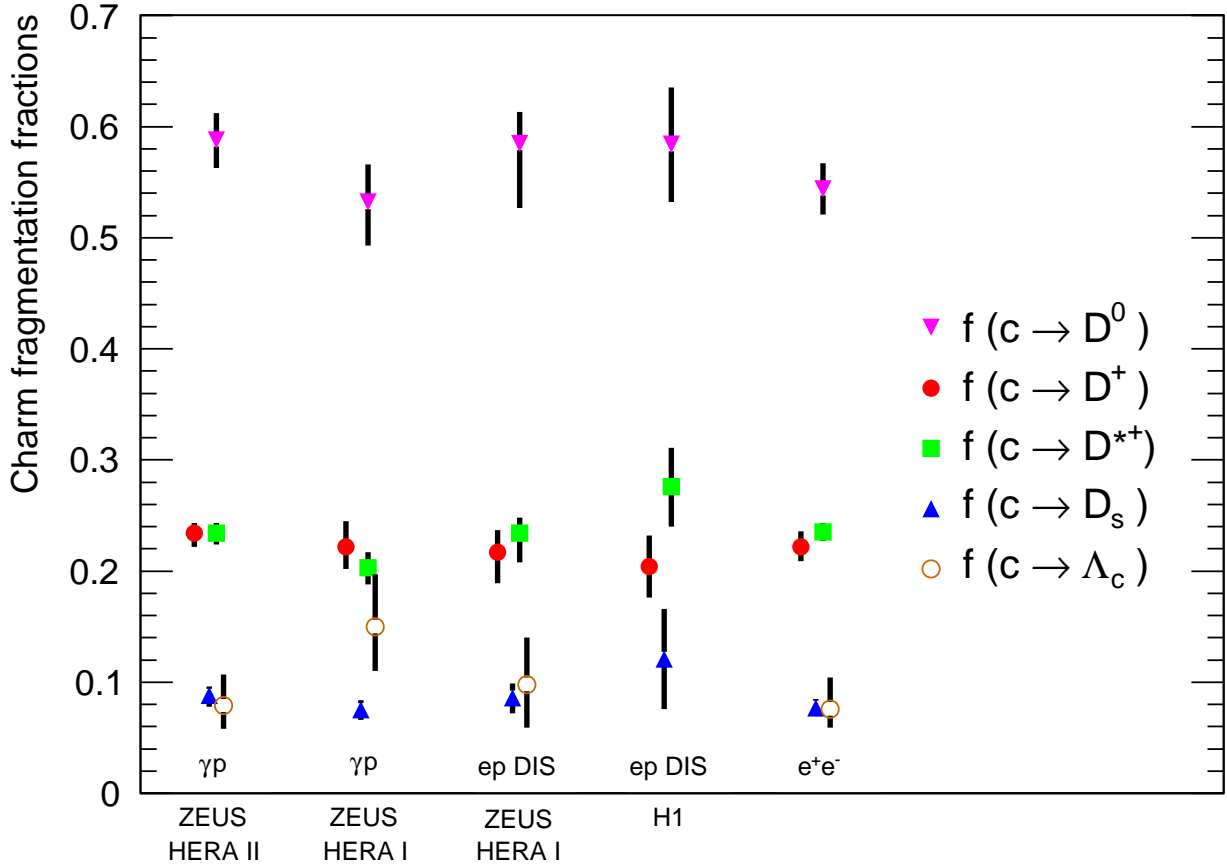}
\end{center}
\caption{Fractions of charm quarks hadronising as a particular charm hadron.
The photoproduction measurements presented in this paper 
are shown (first column) and
compared to previous HERA results in photoproduction (second column),
DIS (third and fourth column) and to $e^{+}e^{-}$ data (last column), 
with statistical, systematic and branching-ratio uncertainties added in 
quadrature.}
\label{1}
\end{figure}
\end{document}